\documentclass[11pt,preprint,preprintnumbers,showpacs,groupedaddress,
superscriptaddress, floatfix,longabstract]{revtex4-1}

\usepackage{amsmath}
\usepackage{amssymb}
\usepackage{bbm}
\usepackage{multirow}
\usepackage{setspace}
\usepackage{graphicx}
\usepackage{epsfig}
\usepackage{dcolumn}
\usepackage{bm}
\usepackage{subfigure}
\usepackage{slashed}
\usepackage{longtable}
\usepackage[colorlinks,
            linkbordercolor={0 0 0},
            pdfborder={0 0 0},
            linkcolor=blue,
            citecolor=blue,
            urlcolor=blue,
            breaklinks=true]{hyperref}

\allowdisplaybreaks

\def\rmsmall #1{\mbox{\scriptsize #1}}
\def\eq#1{\begin{equation} #1 \end{equation}}
\def\eqarray#1{\begin{eqnarray} #1 \end{eqnarray}}

\newcommand{\pol}{\scriptsize\mbox{pol}}
\newcommand{\unpol}{\scriptsize\mbox{unpol}}

\newlength{\x}
\newlength{\y}
\newlength{\z}
\newlength{\leftright}
\begin{document}

\title{A Lattice Study of Quark and Glue Momenta and Angular Momenta in the Nucleon}

\author{M. Deka} %
\email{mpdeka@theor.jinr.ru}%
\affiliation{Bogoliubov Laboratory of Theoretical Physics, JINR, 141980 Dubna, Russia}%
\author{T. Doi}%
\affiliation{Theoretical Research Division, Nishina Center, RIKEN, Wako 351-0198, 
  Japan}%
\author{Y. B. Yang}%
\email{ybyang@pa.uky.edu}%
\affiliation{Institute of High Energy Physics, Chinese Academy of Sciences, 
  Beijing 1000190, China}%
\author{B. Chakraborty}%
\affiliation{\mbox{School of Physics and Astronomy, University of Glasgow, Glasgow, 
  G12 8QQ, UK}}
\author{S. J. Dong}%
\affiliation{\mbox{Department of Physics and Astronomy, University of Kentucky, Lexington,
  KY 40506}}
\author{T. Draper}%
\affiliation{\mbox{Department of Physics and Astronomy, University of Kentucky, Lexington,
  KY 40506}}
\author{M. Glatzmaier}%
\affiliation{\mbox{Department of Physics and Astronomy, University of Kentucky, Lexington,
  KY 40506}}
\author{M. Gong}%
\affiliation{\mbox{Department of Physics and Astronomy, University of Kentucky, Lexington,
  KY 40506}}
\author{H. W. Lin}%
\affiliation{Department of Physics, University of Washington, Seattle, WA 98195}
\author{K. F. Liu}%
\email{liu@pa.uky.edu}%
\affiliation{\mbox{Department of Physics and Astronomy, University of Kentucky, Lexington,
  KY 40506}}%
\author{D. Mankame}%
\affiliation{\mbox{Department of Physics and Astronomy, University of Kentucky, Lexington,
  KY 40506}}%
\author{N. Mathur}%
\affiliation{{Department of Theoretical Physics, Tata Institute of Fundamental 
  Research, Mumbai 40005, India}}%
\author{T. Streuer}%
\affiliation{Institute for Theoretical Physics, University of Regensburg,\ 
    93040 Regensburg, Germany}%

\collaboration{$\chi$QCD Collaboration}\noaffiliation

{\linespread{0.95}
\begin{abstract}
\medskip
%

We report a complete calculation of the quark and glue momenta and angular momenta in 
the proton.\ These include the quark contributions from both the connected and disconnected 
insertions.\ The quark disconnected insertion loops are computed with $Z_4$ noise,\ and the 
signal-to-noise is improved with unbiased subtractions.\ The glue operator is comprised of 
gauge-field tensors constructed from the overlap operator.\ The calculation is carried out on 
a $16^3 \times 24$ quenched lattice at $\beta = 6.0$ for Wilson fermions with $\kappa=0.154,
0.155$,\ and $0.1555$ which correspond to pion masses at $650, 538$,\ and $478$~MeV,\ 
respectively.\ The chirally extrapolated $u$ and $d$ quark momentum/angular momentum 
fraction is found to be $0.64(5)/0.70(5)$,\ the strange momentum/angular momentum fraction 
is $0.024(6)/0.023(7)$,\ and that of the glue is $0.33(6)/0.28(8)$.\ The previous study of
quark spin on the same lattice revealed that it carries a fraction of $0.25(12)$ of proton spin.\ 
The orbital angular momenta of the quarks are then obtained from subtracting the spin from 
their corresponding angular momentum components.\ We find that the quark orbital angular 
momentum constitutes $0.47(13)$ of the proton spin with almost all of it coming from the 
disconnected insertions.

\vspace{1pc}
\end{abstract}
}


\maketitle
\thispagestyle{empty}

\pagestyle{plain}
\setcounter{page}{1}
\pagenumbering{arabic}
\singlespacing
\parskip 0pt
\parindent 10pt


\section{Introduction}

Determining the contributions from quarks and gluons to the nucleon spin is one of the most 
challenging issues in QCD both experimentally and theoretically.\ Since the contribution from 
the quark spin is found out to be small ($\sim$ 25\% of the total proton spin) from the global 
analysis of deep inelastic scattering data~\cite{deFlorian:2009vb},\ it is expected that the 
rest should come from glue spin and the orbital angular momenta of quarks and glue.

\medskip

The quark spin contribution from $u$,~$d$~and~$s$ has been studied on the 
lattice~\cite{Dong:1995rx,Fukugita:1994fh} since 1995 using either the quenched approximation 
or dynamical fermions with heavier quark mass~\cite{Gusken:1999as}.\ Recently,\ it has been 
carried out with light dynamical fermions~\cite{QCDSF:2011aa,Abdel-Rehim:2013wlz},\ and 
only for strange quarks (not renormalized) in~\cite{Babich:2010at}.\ The calculation of 
disconnected insertion (DI) contributions to quark spin from $u$,\ $d$,\ $s$ and 
$c$ using anomalous Ward identity with light overlap fermions is under
progress~\cite{Gong:2013}.

\medskip

As for the quark orbital angular momenta,\ lattice calculations have been carried out for the
connected insertions (CI)~\cite{Mathur:1999uf,Hagler:2003jd,Gockeler:2003jfa,Brommel:2007sb,
  Bratt:2010jn,Alexandrou:2011nr,Alexandrou:2013joa}.\ They are obtained by subtracting 
the quark spin contributions from those of the quark angular momenta.\ It has been shown 
that the contributions from $u$ and $d$ quarks mostly cancel each other.\ Thus for connected 
insertion,\ quark orbital angular momenta turn out to be small in the quenched 
calculation~\cite{Mathur:1999uf,Gockeler:2003jfa} and nearly zero in dynamical fermion 
calculations~\cite{Hagler:2003jd,Brommel:2007sb,Bratt:2010jn,Alexandrou:2011nr,
Alexandrou:2013joa}.\ On the other hand,\ gluon helicity distribution $\Delta G(x)/G(x)$ from 
COMPASS,\ STAR,\ HERMES and PHENIX experiments is found to be close to 
zero~\cite{Adolph:2012ca,Djawotho:2011zz,Airapetian:2010ac,Stolarski:2010zz,Adare:2008qb}.\ 
The latest global fit~\cite{Aschenauer:2013woa} with the inclusion of the polarized deep 
inelastic scattering (DIS) data from COMPASS~\cite{Alekseev:2010hc} and the 2009 data from 
RHIC~\cite{Aschenauer:2013woa},\ gives a glue contribution 
$\displaystyle\int_{0.05}^{0.2} \Delta g(x) dx = 0.1 \pm_{0.07}^{0.06}$ to the total proton spin 
of $1/2 \hbar$ with a sizable uncertainty.\ Furthermore,\ it is argued based on analysis of 
single-spin asymmetry in unpolarized lepton scattering from a transversely polarized nucleon 
that the glue orbital angular momentum is absent~\cite{Brodsky:2006ha}.\ In this given 
context,\ we know from lattice and experiments thus far that $\sim 25\%$ of the proton spin 
comes from the quark spin,\ CI orbital angular momenta have negligible contributions,\ and 
gluon helicity from experiments is $\sim 20\%$.\ Since we have not been able to identify
the rest ($\sim 50\%$) of the proton spin,\ it appears that we have encountered a `Dark Spin' 
conundrum.

\medskip

In this work,\ we give a complete calculation of the quark and glue momenta and angular 
momenta on a quenched lattice.\ The quark contributions to both the connected and 
disconnected insertions are included.\ We have been able to obtain the glue momentum and 
angular momentum for the first time,\ mainly because the overlap operator is used for the 
gauge field tensor~\cite{Liu:2007hq,Alexandru:2008fu},\ the construction of which is much 
less noisy than that from  thin gauge links.\ Combining with earlier work on the quark 
spin~\cite{Dong:1995rx},\ we obtain the quark orbital angular momenta.\ We find that the 
$u$ and $d$ quark orbital angular momenta indeed largely cancel in the connected insertion.\ 
However,\ their contributions together with that of the strange quark are large ($\sim 50$\%) 
in the DI where the quark spin for the $u, d$ and $s$ in DI are large and negative. 

\medskip

These results from our lattice calculations are improved by satisfying the momentum 
and angular momentum sum rules for the quark and glue contributions.\ The renormalization 
and mixing of the quark and glue energy-momentum operators are performed perturbatively,\ 
and the final results are reported in the $\overline{MS}$ scheme at $2$~GeV.

\medskip

The manuscript is organized as follows:\ Section~\ref{formalism} discusses the general 
formalism about the quark and glue energy-momentum tensor operators and their 
contributions to the proton momenta and angular momenta via the associated form factors.\ 
The lattice formulation is presented in Sec.~\ref{lattice}.\ In 
Sec.~\ref{sec:disconnected_insertions},\ the stochastic method for computing the quark 
loops in the disconnected insertions and the computation of glue operators constructed from 
the overlap operator are described.\ Utilization of unbiased subtraction and discrete symmetries 
for variance reduction is also discussed in the same section.\ The choice of momenta and 
the separation of the $T_1, T_2$ and $T_3$ form factors are described in 
Secs.~\ref{sec:choice_of_mom} and \ref{sec:separation_of_t1_t2_t3},\ respectively.\ The 
renormalization of the quark and glue energy momentum tensor operators and their mixing 
and matching to the $\overline{MS}$ scheme at $2$~GeV scale is discussed in 
Sec.~\ref{sec:renormalization}.\ We give the numerical details in 
Sec.~\ref{sec:numerical_param} and the results in Sec.~\ref{sec:results}.\ We conclude 
with a summary in Sec.~\ref{sec:summary}.


\section{General Formalism}    \label{formalism}

The Lorentz group generators,\ $J^{\mu\nu}$,\ for angular momentum operators are given 
by~\cite{Weingerg:1972}
\eqarray{
 J^{\mu\nu} &\equiv& \displaystyle\int d^3 x M^{0\mu\nu} (\vec{x}) .
\label{ang_mom_generator}
} 
Here $M^{0\mu\nu}$ is the angular momentum density which is defined as
\eqarray{
  M^{\mu\nu\alpha} (x) 
  &=& {\mathcal T}^{\{\mu\alpha\}} x^\nu - {\mathcal T}^{\{\mu\nu\}} x^\alpha ,
\label{ang_mom_density}
}
where $\mathcal{T}^{\{\mu\nu\}}$ is the energy-momentum tensor and has the 
Belinfante-improved form.\ It is gauge-invariant and conserved~\cite{Treiman:1986ep},\ 
and $\{\cdots\}$ stands for symmetrization of indices.

\smallskip

\noindent
One can construct gauge-invariant energy-momentum tensor operators for quarks and 
gluons separately.\ As a result,\ we can write  $\mathcal{T}^{\{\mu\nu\}}$ 
as the following gauge-invariant sum
\eqarray{
  \mathcal{T}^{\{\mu\nu\}} 
  &=& \mathcal{T}^{\{\mu\nu\} q} + \mathcal{T}^{\{\mu\nu\} g} ,
  \label{energy_mom_split}
}
where the superscripts,\ $q$ and $g$, stand for quarks and gluons,\ respectively.\ The 
operators,\ $\mathcal{T}^{\{0i\} q}$ and $\mathcal{T}^{\{0i\} g}$,\ have the following 
twist-two forms
\eqarray{
  {\mathcal T}^{\{0i\} q} 
  &=& \frac{i}{4}\displaystyle\sum_f \overline {\psi}_f 
  \left[ 
    \gamma^0 \stackrel{\rightarrow}{D^i}
    + \gamma^i \stackrel{\rightarrow} {D^0}
    -  \gamma^0 \stackrel{\leftarrow} {D^i} 
    - \gamma^i \stackrel{\leftarrow} {D^0}
  \right] \psi_f ,
  \label{q_contrib_def_1}
}
and
\eqarray{
  {\mathcal T}^{\{0i\}g} 
  &=& - \frac{1}{2} \displaystyle\sum_{k=1}^3 
  \left[G^{a, 0k}\, G^{a, i}_k + G^{a,ik}\, G^{a,0}_k \right] 
  = - \frac{1}{2} \displaystyle\sum_{k=1}^3 2\, 
  \mbox{Tr}^{\rmsmall{color}}\left[G^{0k}\, G^{i}_k + G^{ik}\, G^{0}_k \right] .
\label{g_contrib_def_1}
}
In Eq.~(\ref{q_contrib_def_1}),\ $\psi_f$ denotes the quark field operator for the flavor,\ 
$f$.\ In Eq.~(\ref{g_contrib_def_1}),\ $a$ is the color index and $G$'s are the gauge field 
strength tensors.

\

\noindent
Using Eqs.~(\ref{ang_mom_generator}),\ 
(\ref{ang_mom_density})~and~(\ref{energy_mom_split}),\ one can write the $i$-th 
component of the angular momentum operators for quarks and gluons,\ ${\vec J}^{q,g}$,\ 
as
\eqarray{
  J_i^{q,g} 
  &=& \frac{1}{2}\,\epsilon^{ijk}\,\int \, d^3x\, \left(\mathcal{T}^{\{0k\} q,g}\, x^j
    - \mathcal{T}^{\{0j\}q,g}\, x^k\right) ,
\label{ang_op_def_split_1}
}
so that the total angular momentum is
\eqarray{
\vec{J} &=& \vec{J}^q + \vec{J}^g .
\label{tot_ang_mom_sum}
}

\noindent
In a similar manner,\ the linear momentum operators are given by
\eqarray{
  P_i^{q,g} &=& \int \, d^3x\, \mathcal{T}^{\{0i\} q,g} .
\label{momentum_op}
}
Substituting the explicit form of ${\mathcal T}^{\{0i\} q}$ from Eq.~(\ref{q_contrib_def_1}) 
into Eq.~(\ref{ang_op_def_split_1}),\ and using the QCD equations of motion,\ one can 
obtain the gauge-invariant decomposition of $\vec{J}^q$ as~\cite{Jaffe:1989jz,ji97}
\eqarray{
  \vec{J}^q &=& 
  \int d^3x \, 
  \bigg{[} \frac{1}{2}\, \overline\psi\,\vec{\gamma}\,\gamma^5 \,\psi 
 + \psi^\dag \,\{ \vec{x} \times (i \vec{D}) \} \,\psi \bigg{]} ,
\label{quark_ang_op_split_1}
}
where the color indices are suppressed.\ From the spin-$\frac{1}{2}$ field theory,\ one can 
identify the first term of Eq.~(\ref{quark_ang_op_split_1}) as the quark spin operator 
($\displaystyle\frac{1}{2}\vec\Sigma^q$) and the second term as the orbital angular 
momentum operator ($\vec{L}^q$).\ Thus,\ we can write the total angular momentum for 
quarks as
\eqarray{
\vec{J}^q &=&  \frac{1}{2} \vec\Sigma^q + \vec{L}^q .
\label{quark_ang_op_def_split_2}
}
Similarly, using equations of motion and superpotentials, it is shown that 
Eqs.~(\ref{g_contrib_def_1})~and~(\ref{ang_op_def_split_1}) lead to a gauge invariant 
glue angular momentum operator~\cite{Jaffe:1989jz,ji97} as
\eqarray{
\vec{J}^g &=& \int d^3x \,\bigg{[} \vec{x} \times ( \vec{E} \times \vec{B} )\bigg{]} ,
\label{gluon_ang_op_def_split_1}
}
where $\vec{E}$ and $\vec{B}$ are the electric and magnetic fields for gluons,\ respectively.\ 
Hence,\ the angular momentum operator in QCD can be expressed as the following 
gauge-invariant sum of operators~\cite{ji97}
\eq{
  \vec J_{\rmsmall{QCD}} = \vec J^q + \vec J^g
  = \frac{1}{2} \vec\Sigma^q + \vec{L}^q + \vec{J}^g .
\label{ang_op_def_split_2}
}
They represent the quark spin, the quark orbital angular momentum and glue angular
momentum, respectively.\ There are discussions in the literature as to whether the glue
operator can be further decomposed into the spin and orbital angular momentum as
in the case of the quarks~\cite{Jaffe:1989jz,Chen:2008ag,Wakamatsu:2010qj,Hatta:2011zs,
  Ji:2013fga, Leader:2013jra}.\ We shall not address this issue in the present work.

\

In order to identify the missing `Dark Spin' from first principles,\ we need to measure 
all the three quantities in Eq.~(\ref{ang_op_def_split_2}) using Lattice QCD.

\

From the first term of the Eq.~(\ref{quark_ang_op_split_1}),\ we see that the spin 
contribution from quarks can be computed using the flavor singlet axial-vector operator,\ 
$\overline\psi\, \gamma^\mu\,\gamma^5\,\psi$,\ and it has a well-defined matrix element.\ 
There have already been a few studies on the lattice~\cite{Dong:1995rx,Fukugita:1994fh,
  Gusken:1999as,QCDSF:2011aa} in this regard.\ However,\ both the second term of the 
Eq.~(\ref{quark_ang_op_split_1}) and the term of the Eq.~(\ref{gluon_ang_op_def_split_1}) 
involve the spatial coordinate $\vec{x}$.\ While they are natural operators for hadronic 
models where the origin of the proton is prescribed,\ it is shown that a straight-forward 
application of the lattice calculation of the moments of the spatial coordinate is complicated 
by the periodic condition of the lattice,\ and will lead to wrong results~\cite{Wilcox:2002zt}.\ 
Hence,\ instead of calculating the orbital angular momentum $L^q$ directly using lattice,\ we 
will calculate the total angular momentum $J^q$ for quarks,\ and then subtract the quark 
spin contributions to determine $L^q$. 

\smallskip

The matrix element of ${\mathcal T}^{(0i) q,g}$ between two nucleon states can be 
written in terms of three form factors ($T_1, T_2$ and $T_3$) as~\cite{ji97}
\eqarray{
(p',s' | {\mathcal T}^{\{0i\} q,g} | p,s)
  &=& \left(\frac{1}{2}\right) \bar{u}(p',s') \left[T_1(q^2)(\gamma^0\bar{p}^i 
   +  \gamma^i\bar{p}^0) 
   + \frac{1}{2m}T_2(q^2)\left(\bar{p}^0 (i \sigma^{i\alpha})
   +  \bar{p}^i (i \sigma^{0\alpha})\right) q_{\alpha}\right.\nonumber\\
  &+& \left.\frac{1}{m} T_3(q^2) q^0 q^i\right]^{q,g} u(p,s) ,
\label{mat_element_1}
}
where $p$ and $p'$ are the initial and final momenta of the nucleon,\ respectively,\ and 
$\bar{p} = \displaystyle\frac{1}{2}\, (p' + p)$.\ $q_\mu = p'_\mu - p_\mu$ is the momentum 
transfer to the nucleon,\ $m$ is the mass of the nucleon, and $u(p,s)$ is the nucleon spinor.\ 
$s$ and $s'$ are the initial and final spins,\ respectively.\ The spinor,\ $u(p,s)$,\ satisfies the 
following normalization conditions
\eq{
\bar{u}(p,s)\, u(p,s)\, =\, 2m\, , \, 
\displaystyle\sum_s  u(p,s)\, \bar{u}(p,s)\, = \, \slashed{p} + m .
\label{norm_cond_min}
}

\smallskip

\noindent
By substituting Eq.~(\ref{mat_element_1}) into Eqs.~(\ref{ang_op_def_split_1}) 
and (\ref{momentum_op}),\ and then taking  $q^2 \rightarrow 0$ limit,\ one obtains
\eqarray{
  \label{ang_op_def_split_3}
  J^{q,g} &=& \frac{1}{2} \left[T_1(0) + T_2(0)\right]^{q,g} ,\\
  \langle x\rangle^{q,g} &=& T_1(0)^{q,g} .
\label{momentum_fraction}
}
where $\langle x\rangle^{q,g} = T_1(0)^{q,g}$ is the first moment of the momentum 
fraction carried by the quarks or gluons inside a nucleon.\ The other form factor,\ 
$T_2(0)^{q,g}$,\ can be interpreted as anomalous gravitomagnetic moment for quarks 
and gluons in an analogy to the anomalous magnetic moment,\ 
$F_2(0)$~\cite{Teryaev:1999su}.

\smallskip

\noindent
Since momentum is always conserved and the nucleon has a total spin of 
$\displaystyle\frac{1}{2}$,\ we write the momentum and angular momentum sum rules 
using 
Eqs.~(\ref{ang_op_def_split_2}),~(\ref{ang_op_def_split_3})~and~(\ref{momentum_fraction}),\ 
as
\eqarray{
  \label{eq:mom_sum_rule}
  \langle x\rangle^{q} +   \langle x\rangle^{g}\, =\, T_1 (0)^q + T_1 (0)^g  &=& 1 , \\
  J^q+ J^g\, =\, \frac{1}{2}\,  \bigg\{
  \left[ T_1 (0) + T_2 (0) \right]^q + \left[ T_1 (0)  + T_2 (0) \right]^g 
  \bigg\} &=& \frac{1}{2} . 
  \label{eq:ang_mom_sum_rule}
}
It is interesting to note that from Eqs.~(\ref{eq:mom_sum_rule}) and 
(\ref{eq:ang_mom_sum_rule}),\ one obtains that the sum of the $T_2(0)$'s for the 
quarks and glue is zero,\ i.e.
\eqarray{
T_2 (0)^q + T_2 (0)^g &=& 0 .
\label{eq:T_2_sum}
}
The vanishing of $T_2 (0)$ in the context of a spin-$1/2$ particle was first derived classically from 
the post-Newtonian manifestation of equivalence principle~\cite{Kobzarev:1962wt}.\ More 
recently,\ this has been proven by Brodsky {\it et al.}~\cite{Brodsky:2000ii} for composite 
systems from the light-cone Fock space representation.

\smallskip

Since we are going to evaluate $J^{q,g}$ (or $L^{q,g}$) and $\langle x\rangle^{q,g}$ in this 
work,\ it is clear from Eqs.~(\ref{ang_op_def_split_3})~and~(\ref{momentum_fraction}) 
that we need to compute both $T_1(0)$ and $T_2(0)$.\ However,\  $T_2(0)$  can not be 
computed  directly at $q = 0$ because the $T_2$ form factor in Eq.~(\ref{mat_element_1}) is 
proportional to $q$.\ Instead,\ we shall compute $T_1(q^2)$ and $T_2(q^2)$ separately 
at some $q^2 \neq 0$ values~\cite{Hagler:2003jd}, and then separately extrapolate them to 
$q^2 \rightarrow 0$ for both the quark and glue contributions.


\section{Lattice Formalism}   \label{lattice}

\subsection{Operators and Matrix Elements in Euclidean Space-time}

In order to carry out lattice calculations,\ we use the Pauli-Sakurai convention~\cite{montvay,
Best:1997qp,Deka:2008xr} for the $\gamma$ matrices in Euclidean space-time.\ We 
can then write the energy momentum tensor for quarks and gluons as
\eqarray{
\label{eq:q_contrib_def_2}
  {\mathcal T}_{\{4i\}}^{q (E)} 
  &=&  (-1)\, \frac{i}{4}\displaystyle\sum_f \overline {\psi}_f 
  \left[ 
    \gamma_4 \stackrel{\rightarrow} D_i 
    + \gamma_i \stackrel{\rightarrow} D_4
    -  \gamma_4 \stackrel{\leftarrow} D_i
    - \gamma_i \stackrel{\leftarrow} D_4
    \right] \psi_f , \\
\label{eq:g_contrib_def_2}
  {\mathcal T}_{\{4i\}}^{g (E)} 
  &=&  (+i)\, \bigg[-\frac{1}{2} \displaystyle\sum_{k=1}^3 2\, 
    \mbox{Tr}^{\rmsmall{color}} \left[G_{4k}\, G_{ki} + G_{ik}\, G_{k4} \right]\bigg] .
}
The matrix elements for both quarks and gluons transform in a similar manner as
\eqarray{
\langle p', s'  | {\mathcal T}_{\{4i\}}^{q,g (E)} | p, s\rangle
  &=& \left(\frac{1}{2}\right) \bar{u}^{(E)}(p',s') \left[T_1(-q^2)(\gamma_4\bar{p}_i 
   +  \gamma_i\bar{p}_4) 
   - \frac{1}{2m}T_2(-q^2)(\bar{p}_4 \sigma_{i\alpha} q_{\alpha} 
   +  \bar{p}_i \sigma_{4\alpha} q_{\alpha})\right.\nonumber\\
  &-& \left.\frac{i}{m} T_3(-q^2) q_4 q_i\right]^{q,g} u^{(E)}(p,s) .
\label{mat_element_2}
}
where the normalization conditions are
\eq{
\bar{u}^{(E)}(p,s)\, u^{(E)}(p,s)\, =\, 1\, , \, 
\displaystyle\sum_s  u^{(E)}(p,s)\, \bar{u}^{(E)}(p,s)\, 
= \, \frac{\slashed{p} + m}{2m} , 
\label{norm_cond_euc}
}
and the L.H.S. of Eqs.~(\ref{mat_element_1}) and (\ref{mat_element_2}) are related 
by
\eq{
  \displaystyle\frac{(p',s'}{\sqrt{2m}} 
  \left| {\mathcal T}^{\{0i\} q,g} 
  \right | \frac{p,s)} {\sqrt{2m}}\,
  \longleftrightarrow\,
  \langle p', s' |\, {\mathcal T}_{\{4i\}}^{q,g (E)}\, | p, s\rangle
  \label{mat_elem_lhs_min_2_euc}
}
From now on,\ we will consider the Euclidean operators only and drop the superscript,\ $E$.

\subsection{Quark Energy-momentum Tensor Operator}
\label{subsec:lat_quark_ang_mom_op}

We discretize ${\mathcal T}_{\{4i\}}^{q}$ by using the following relations for right and
left derivatives in lattice~\cite{Kronfeld:1984zv}
\eqarray{
  \label{eq:discretn1}
  \stackrel{\rightarrow} D_\mu \psi^L(x) &=& \frac{1}{2a}\,
  \left[ U_\mu (x)\, \psi^L (x + a_\mu) - U^\dag_\mu(x-a_\mu)\, \psi^L(x-a_\mu) \right] ,\\
  \overline{\psi}^L(x)\stackrel{\leftarrow} D_\mu &=&
  \frac{1}{2a}\, \left[ \overline\psi^L (x + a_\mu)\, U^\dag_\mu (x) -
    \overline\psi^L(x - a_\mu)\, U_\mu(x-a_\mu) \right] ,
  \label{eq:discretn2}
}
where $a$ is the lattice spacing,\ $\psi^L$'s are the Lattice quark field operators,\ and $U$'s 
are the gauge links.\ Using the relations in 
Eqs.~(\ref{eq:discretn1})~and~(\ref{eq:discretn2}),\ we get
\eqarray{
  {\mathcal T}_{\{4i\}}^q (x) 
  &=& \frac{-i}{8a}\,\left[ \overline\psi_{f}(x)\,\gamma_4\,
    U_i (x) \, \psi_f(x + a_i)\, - \, \overline\psi_f(x)\,\gamma_4\,
    U^\dag_i (x - a_i) \, \psi_f(x - a_i)\right. \nonumber\\
  & &\nonumber\\
  &+& \overline\psi_f(x - a_i)\,\gamma_4\, U_i (x - a_i) \, \psi_f(x) -
  \overline\psi_f(x + a_i)\,\gamma_4\, U^\dag_i (x) \, \psi_f(x)\nonumber\\
  & &\nonumber\\
  &+& \overline\psi_f(x)\,\gamma_i\, U_4 (x) \, \psi_f(x + a_4) -
  \overline\psi_f(x)\,\gamma_i\, U^\dag_4 (x - a_4) \, \psi_f(x - a_4)\nonumber\\
  & &\nonumber\\
  &+& \left. \overline\psi_f(x - a_4)\,\gamma_i\, U_4 (x - a_4) \, \psi_f(x) -
    \overline\psi_f(x + a_4)\,\gamma_i\, U^\dag_4 (x) \, \psi_f(x) \right] .
  \label{eq:quark_op_discrete}
}
%

\subsection{Glue energy-momentum tensor operator}  
\label{subsec:lat_field_tensor_op}

It is well known that gauge operators obtained from the link variables are very noisy due to 
the large fluctuations in high-frequency modes.\ A preliminary study of the glue momentum 
fraction in the nucleon on a quenched lattice concluded that configurations in the order of 
several hundred thousands might be needed for a precise signal~\cite{Gockeler:1996zg}.\ 
This is a tall order for dynamical fermion calculations.\ On the other hand,\ a smeared 
operator will improve the signal,\ and HYP smearing has been applied to calculate the glue 
momentum fraction in the pion with reasonable precision~\cite{Meyer:2007tm}. 

\smallskip

Due to the exponentially local nature of the overlap Dirac operator through chiral 
smearing~\cite{Hernandez:1998et,Draper:2005mh,Draper:2006wb},\ the subdimensional 
long range order of the topological structure has been discovered~\cite{Horvath:2003yj,
  Horvath:2005cv} with the help of the local topological charge operator derived from the 
massless overlap Dirac operator,\ i.e. 
$q(x) = \mbox{Tr}\, \gamma_5 (1 - \frac{1}{2} D_{\rmsmall{ov}} (x,x))$~\cite{Kikukawa:1998pd,
Adams:1998eg,Fujikawa:1998if,Suzuki:1998yz}.\ Prompted by this success,\ it is shown that 
the gauge field tensor can be similarly derived from the massless overlap operator 
$D_{\rmsmall{ov}}$~\cite{Liu:2007hq, Alexandru:2008fu}
\eqarray{
  \mbox{Tr}_s \left[\sigma_{\mu\nu} D_{\rmsmall{ov}}(x,x)\right] 
  &=&  c_T\, a^2\, G_{\mu\nu}(x) + {\mathcal O}(a^3) ,
\label{Glue_Tensor}
}
where $\mbox{Tr}_s$ is the trace over spin.\ $c_T = 0.11157$ is the proportional constant 
at the continuum limit for the parameter $\kappa = 0.19$ in the Wilson kernel of the overlap 
operator which is used in this work.\ The glue energy momentum tensor in 
Eq.~(\ref{eq:g_contrib_def_2}) constructed with this gauge field tensor was used in 
calculating the glue momentum fraction,\ $\langle x \rangle_g$,\ which resulted in a first 
observation with a much better signal~\cite{Doi:2008hp} .

\smallskip

We shall use the energy momentum tensor with the noise-estimated gauge field tensor as
defined in Eq.~(\ref{eq:g_contrib_def_2}) from the overlap Dirac operator to calculate the
glue momentum and angular momentum in the nucleon.

\subsection{Two-Point Correlation Functions}
\label{subsec:twopoint_funcs}

In order to obtain $T_1(0)^{q,g}$ and $T_2(0)^{q,g}$,\ we first need to calculate the two-point 
and three-point functions (both polarized and unpolarized) for protons (neutrons).\ The 
two-point function is defined (with the color indices suppressed) as
\eqarray{
  G_{\alpha\beta}^{NN}(\vec{p}, t; t_0) 
  &=&\displaystyle\sum_{\vec{x}}\, e^{-i\vec{p} \cdot (\vec{x}-\vec{x}_0)}\,
  \langle 0 |
  \mbox{T} \left[\chi_\alpha(\vec{x}, t)\, \bar{\chi}_\beta(\vec{x}_0, t_0)\right] 
  | 0 \rangle ,
  \label{eq:two_point_func}
}
where $t$ is the nucleon sink time,\ $\vec{p}$  is the momentum of the nucleon,\ and ${x_0}$ 
is the source position.\ The interpolating fields,\ $\chi$'s,\ for nucleons that we use are given 
by~\cite{Ioffe:1981kw,Draper:1989pi}
\eqarray{
  \label{eq:interpol_fileds_1}
  \chi_\gamma(x)
  &=&\epsilon_{abc}\,\psi\,^{T (u)a}_\alpha(x)\,
  (C \gamma_5)_{\alpha\beta} \, \psi^{(d)b}_\beta(x)\, \psi^{(u)c}_\gamma(x), \\
  %
  %
  \label{eq:interpol_fileds_2}
  \bar{\chi}_{\gamma^\prime} (x)
  &=& -\epsilon_{def}\,\overline{\psi}^{(u)f}_{\gamma^\prime}(x)\,
  \overline{\psi}^{(d)e}_{\rho}(x)\,(\gamma_5 C)_{\rho\sigma}\,
  \overline{\psi}\,^{T (u)d}_{\sigma}(x),
}
where $u$ and $d$ stand for up and down quarks,\ respectively.\ $C=\gamma_2\gamma_4$ 
is the charge conjugation operator with the Pauli-Sakurai $\gamma$ matrices.\ The letters,\
$a$,\ $b$,\ $\cdots$,\ stand for the color indices.\ The Greek letters,\ $\alpha$,\ $ \beta$,\
$\cdots$,\ are the spin indices.

\smallskip

\noindent
Upon Grassmann integration for Eq.~(\ref{eq:two_point_func}),\ we obtain the 
unpolarized/polarized proton two-point function on a gauge configuration $U$ as
\eqarray{
  &  &\mbox{Tr}\left[\Gamma^{\unpol,\pol}\, G^{NN} (\vec{p}, t; t_0; U)\right]
  \nonumber\\ 
  &=&\displaystyle\sum_{\vec{x}} e^{- i\vec{p} \cdot (\vec{x} - \vec{x}_0)}
  \epsilon_{abc}\,\epsilon_{def}\nonumber\\ 
  & & \bigg{\lbrace}
  \mbox{Tr} \left[\Gamma^{\unpol,\pol} \, S^{(u){ad}}(x,x_0;U)\right]\,
  \mbox{Tr}\left[\tilde{S}^{(u){be}}(x,x_0;U)\,S^{(d){cf}}(x,x_0;U)\right] 
  \nonumber\\
  & & + \mbox{Tr}\left[\Gamma^{\unpol,\pol}\, S^{(u){ad}}(x,x_0;U)\,
    \tilde{S}^{(d){be}}(x,x_0;U)\,S^{(u){cf}}(x,x_0;U)\right]\bigg{\rbrace}\nonumber\\
  &=& \displaystyle\sum_{\vec{x}} e^{-i\vec{p} \cdot (\vec{x}  -\vec{x}_0)}\, 
  N^{\unpol,\pol} [x;U] ,
}
where $\Gamma^{\unpol,\pol}$ are the unpolarized/polarized projection operators,\ and 
$N^{\unpol,\pol} [x; U]$ stands for the trace part of two-point functions ({\it sans} the
Fourier factor).\ $S^{(f)}(x, y;U)$ is the quark propagator with flavor~$f$ from the 
point $y$ to $x$ on the gauge configuration $U$,\ and 
\eq{
\tilde S = (C\, \gamma_5)^{-1}\, S^T\, (C\, \gamma_5) ,
}

\smallskip

\noindent
On the other hand,\ if we insert a complete set of energy eigenstates in 
Eq.~(\ref{eq:two_point_func}),\ and take the trace with the unpolarized projection 
operator,\ then at a large time separation we get the two-point functions for nucleons
as
\eqarray{
  \mbox{Tr}\left[\Gamma^{\unpol}\, G^{NN} (\vec{p}, t; t_0)\right]
  &\hspace{2mm}
  {\overrightarrow{\hspace{\x}}}\hspace{-\x}\raisebox{2ex}{$(t-t_0)\gg 1/\Delta E$}
  \hspace{3mm}& 
  \frac{a^6}{(2k)^3}\, |\phi (p)|^2\, e^{-i\vec{p}.\vec{x}_0}\, \frac{E_p + m}{E_p}\, 
  e^{-E_p (t-t_0)} ,
\label{eq:unpol_two_point_func_1}
}
where $\kappa$ is the hopping parameter,\ $m$ is the mass of nucleon,\ and $E_p$ is its 
ground state energy.\ $\Delta E$ is the energy gap between the ground state and first excited 
state.\ $\phi(p)$ is the vacuum to nucleon transition matrix element due to the 
interpolation field $\chi$,\ and we treat it as a function of $p$ to account for the possible
$p$-dependent lattice systematics.

\subsection{Three-Point Correlation Functions}
\label{subsec:threepoint_funcs}

The three-point functions for ${\mathcal T}_{\{4i\}}^{q,g}$ (or,\ any generic operator) is 
defined as
\eqarray{
  G_{\alpha\beta}^{N {\mathcal T}_{4i} N}(\vec{p}\,',t_2;\vec{q},t_1; \vec{p},t_0)
  &=& \displaystyle\sum_{\vec{x}_1,\vec{x}_2}\, 
  e^{-i\vec{p}\,' \cdot (\vec{x}_2-\vec{x}_1)}\,  
  e^{-i\vec{p} \cdot (\vec{x}_1-\vec{x}_0)}\nonumber\\
  &\times& \langle 0 |
  \mbox{T} \left[\chi_{\alpha}(\vec{x}_2,t_2)\, {\mathcal T}_{\{4i\}} (\vec{x}_1,t_1)\,  
    \bar{\chi}_{\beta}(\vec{x}_0,t_0)\right] | 0 \rangle ,
  \label{eq:three_point_func_1}
}
where $t_2$ is the nucleon sink time,\ $t_1$ is the current insertion time,\ $t_0$ is the 
nucleon source time.\ $\vec{p}$  and $\vec{p}\,'$ are the initial and final momenta of the 
nucleon,\ respectively,\ and $\vec{q} = \vec{p}\,'  - \vec{p}$ is the momentum transfer.

\smallskip

\noindent
By inserting a complete set of energy eigenstates in Eq.~(\ref{eq:three_point_func_1})
and then taking the trace with $\Gamma^{\unpol,\pol}$,\ we get the three-point functions 
as
\eqarray{
  &  &\mbox{Tr}\big{[}\Gamma^{\unpol,\pol}\, G^{(f){N {\mathcal T}_{4i} N}}
  (\vec{p}\,',t_2;\vec{q},t_1; \vec{p},t_0)\big{]} \hspace{3mm}
  {\overrightarrow{\hspace{\z}}}\hspace{-\z}\raisebox{2ex}
  {$(t_1 - t_0),\, (t_2 - t_1) \gg 1/\Delta E$}\nonumber \\
  &  &
  \frac{a^6}{(2\kappa)^3}\, \phi (p)\, \phi (p')\, e^{-i \vec{p}.\vec{x}_0}\frac{1}{4}\, 
  \frac{1}{2\kappa}\, \frac{1}{E_p E_{p'}}\, 
  e^{-E_p (t_2-t_1)}\, e^{-E_{p'} (t_1-t_0)} 
  \left[a_1\, T_1(q^2) + a_2\, T_2(q^2) + a_3\, T_3(q^2)\right] , \nonumber \\
  }
where $a_i$'s are constant coefficients which depend upon the momentum and energy of the 
proton and,\ therefore,\ are known {\it a priori}.\ For the unpolarized case,\ they can be 
written as
\eqarray{
  a_1 
  &=& (p'_i + p_i) ((E_p + m) (E_{p'} + m) + p'_j p_j)
  + (E_{p'} + E_p) (p'_i (E_p + m) + p_i (E_{p'} + m)) ,\nonumber\\
  a_2 
  &=& - \frac{1}{2m}\, \left\{(E_p + m) ((E^2_{p'} - E^2_p) p'_i + (p'_i + p_i) q_j p'_j)
  \right. \nonumber\\
  &-& \left. (E_{p'} + m) ((E^2_{p'} - E^2_p) p_i -  (p'_i + p_i) q_j p_j) 
  +  (E_{p'} + E_p) (p'_j p_i q_j - p'_i p_j q_j)\right\} ,\nonumber\\
  a_3 
  &=& \frac{2}{m}\, (E_{p'} - E_p) q_i ((E_p + m)(E_{p'} + m) - p'_j p_j) ,
  \label{unpol_coeff}
} 
and for the polarized case they are
\eqarray{
  a_1 
  &=& (-i)\, \epsilon_{ijl} (E_{p'} + E_p) (p'_j (E_p + m) - p_j (E_{p'} + m)) 
  +\, \epsilon_{kjl} (p'_i + p_i) p'_j p_k ,\nonumber\\
  a_2 
  &=&  \frac{-i}{2m}
  \left\{(E_p + m)(E_{p'} + m)\epsilon_{ijl} (E_{p'} + E_p) q_j \right.\nonumber\\
  &-& ((E_p + m) p'_j + (E_{p'} + m) p_j) (\epsilon_{ijl} (E^2_{p'} - E^2_p)
  + \epsilon_{kjl} (p'_i + p_i) q_k) \nonumber\\
  &-& \left. (\epsilon_{ijk} (E_{p'} + E_p) (p'_l p_k q_j + p'_j p_l q_k)
  + \epsilon_{ijl} (E_{p'} + E_p) p'_k p_k q_j)\right\} ,\nonumber\\
  a_3
  &=&  \frac{- 2 i}{m}\, \epsilon_{kjl} (E_{p'} - E_p) q_i p'_j p_k .
\label{pol_coeff}
}
Here the subscript $i$ stands for the spatial direction of the energy-momentum operator,\ and 
$l$ is the direction of polarization of the nucleon.
\begin{figure}[h]
\centering
\subfigure[]
{\rotatebox{0}{\includegraphics[width=0.44\hsize]{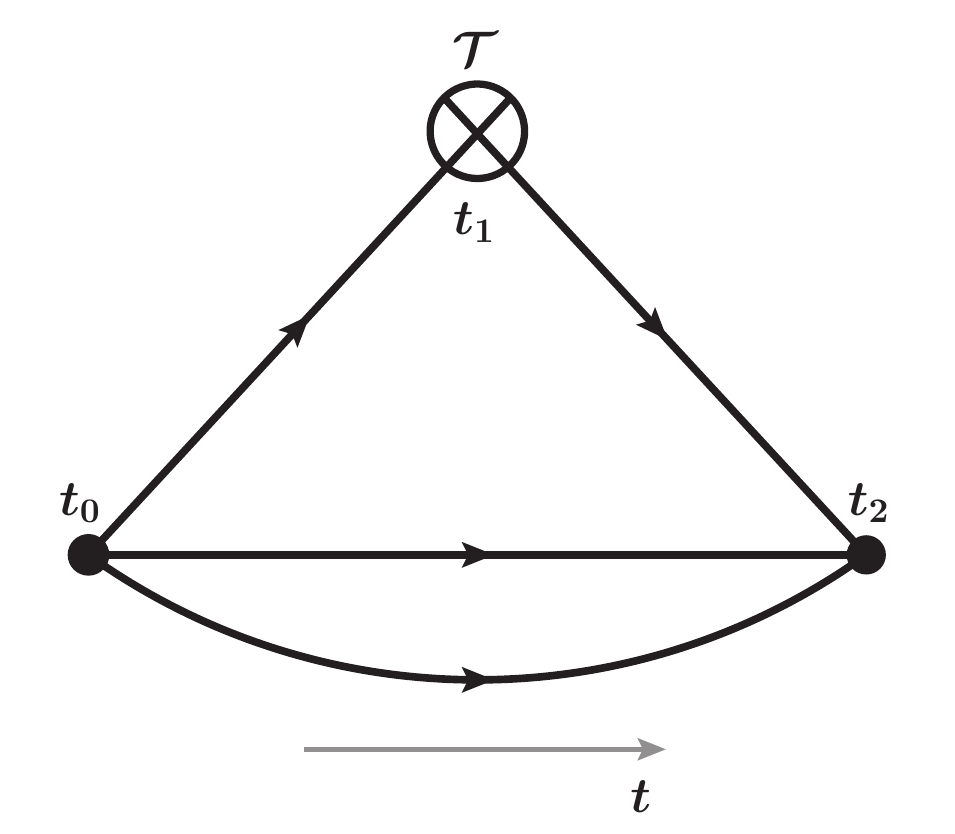}}
\label{fig:connected_insertion}}
\hspace{2mm}
\subfigure[]
{\rotatebox{0}{\includegraphics[width=0.44\hsize]{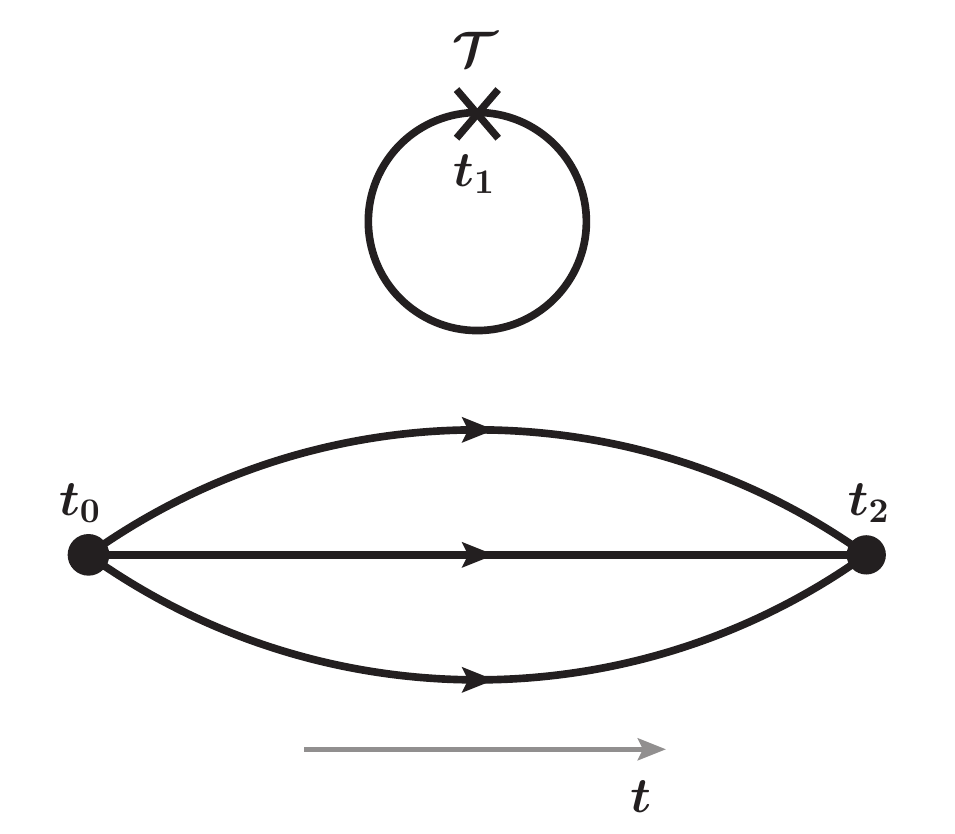}}
 \label{fig:disconnected_insertion}}
\caption{Quark line diagrams of the three-point function with current insertion in the Euclidean
         path integral formalism.\ 
         (a) Connected insertions (CI), and
         (b) disconnected insertions (DI).}
\label{fig:ci_and_di}
\end{figure}

\smallskip

The three-point functions for quarks have two topologically distinct contributions in the
path-integral diagrams:\ one from connected insertions (CI) and the other from 
disconnected insertions (DI)~\cite{Liu:1993cv,Liu:1998um,Liu:1999ak} (See 
Figs.~(\ref{fig:ci_and_di})).\ They arise purely out of Wick contractions,\ and 
it needs to be stressed that they are not Feynman diagrams in perturbation theory.\ In 
the case of CI,\ quark/anti-quark fields from the operator are contracted with the 
quark/anti-quark fields of the proton interpolating fields.\ In the case of DI,\ the 
quark/anti-quark fields from the operator contract themselves to form a current loop,\ as 
in the case of vacuum polarization.

\smallskip

Though not shown in the figure,\ the loop is in fact connected with the proton propagator
through the gauge background fluctuations.\ In practice,\ the uncorrelated part of
the loop and the proton propagator is subtracted.\ The disconnected insertion 
refers to the fact that the quark lines are disconnected. 

\smallskip

For currents with up and down quarks we have contributions from both CI and DI,\ and 
with strange quarks we have DI only.

\smallskip

The current loop for our operator~${\mathcal T}_{\{4i\}}$ with a quark flavor $f$ is given 
by
\eqarray{
  & & L [t_1,\vec{q};U]\nonumber\\
  &=& (-1)\, \frac{- i}{8a} 
  \displaystyle\sum_{\vec{x}_1} e^{i\vec{q} \cdot (\vec{x}_1 - \vec{x}_{0})}\nonumber\\
  &\bigg{\{}&\mbox{Tr}\big{[}S^{(f)mn}(x_1 + a_i, x_1;U)\, \gamma_4\,
  U^{n m}_i(x_1)\big{]} 
  - \mbox{Tr}\big{[}S^{(f)mn}(x_1 - a_i, x_1;U)\,\gamma_4\,U^{\dag n m}_i(x_1 - a_i)\big{]}
  \nonumber\\
  &+& \mbox{Tr}\big{[}S^{(f)mn}(x_1, x_1 - a_i)\,\gamma_4\,U^{n m}_i(x_1-a_i)\big{]} 
  - \mbox{Tr}\big{[}S^{(f)mn}(x_1, x_1 + a_i;U)\,\gamma_4\,U^{\dag n m}_i(x_1)\big{]}
  \nonumber\\ 
  &+& \mbox{Tr}\big{[}S^{(f)mn}(x_1 + a_4, x_1;U)\,\gamma_i\,U^{n m}_4(x_1)\big{]}
  - \mbox{Tr}\big{[}S^{(f)mn}(x_1 - a_4, x_1;U)\,\gamma_i\,U^{\dag n m}_4(x_1 - a_4)\big{]}
  \nonumber\\
  &+&\mbox{Tr}\big{[}S^{(f)mn}(x_1,x_1 - a_4;U)\,\gamma_i\,U^{nm}_4(x_1-a_4)\big{]}
  - \mbox{Tr}\big{[}S^{(f)mn}(x_1, x_1 + a_4;U)\,\gamma_i\,U^{\dag n m}_4(x_1)
  \bigg{\rbrace}\nonumber\\
  &=& \frac{+ i}{8a} \displaystyle\sum_{\vec{x}_1} e^{i\vec{q} \cdot 
    (\vec{x}_1 - \vec{x}_{0})}\, L [x_1; U] ,
  \label{eq:disconnected_4i_f_1}
}
where $L [x_1; U]$ is the trace part of the current loop.\ The gauge-averaged 
DI three-point functions can then be written as
\eqarray{
  & & \mbox{Tr}\big{[}\Gamma^{\unpol,\pol}\, G^{(f) {N {\mathcal T}_{4i} N}}
  (\vec{p}\,',t_2;\vec{q},t_1; \vec{p},t_0)\big{]}_{\rmsmall{DI}}\nonumber\\
  &=& \langle \mbox{Tr}\left[\Gamma^{\unpol,\pol}\, G^{NN} (\vec{p}\,', t_2; t_0;U)\right] 
  \times L [t_1,\vec{q};U] \rangle \nonumber\\
  &-& \langle \mbox{Tr}\left[\Gamma^{\unpol,\pol}\, G^{NN} 
    (\vec{p}\,', t_2; t_0;U)\right] \rangle \times \langle L [t_1,\vec{q};U] \rangle ,
  \label{eq:disconnected_4i_f_2}
}
where $\langle\cdots\rangle$ denotes the average over the gauge ensemble.\ It is 
important to note that the three-point functions for gluons have the similar form 
as DI.

\smallskip

The computation of the CI is relatively straightforward.\ We shall use the sequential 
source technique~\cite{bernard_1,Draper:1984xz,Bernard:1985tm} for CI.\ This fixes 
the source point $t_0$ and the sink time slice $t_2$.\ However,\ the computation of the 
DI is numerically challenging as it contains not only the usual propagators from the 
source,\ $x_0$,\ to any point,\ $x$,\ but also the propagators from any insertion position 
($x_1$) to any other lattice points.\ This would require inverting the fermion matrix at each 
point of the lattice to construct the {\em all-to-all} propagators.\ Naively,\ this entails 
inversion of a million by million ($\sim16^3\times24\times3\times4$) sparse matrix for 
our $16^3\times24$ lattice ($3$~and~$4$~being the number of color and spin indices,\ 
respectively) on each gauge configuration.\ This is unattainable by using the computing 
powers of today's supercomputers.\ Instead,\ we shall compute it with the stochastic 
method.\ Specifically,\ we adopt the complex $Z_2$ noise~\cite{Dong:1993pk} for the 
estimation together with unbiased subtraction~\cite{Thron:1997iy} to reduce variance.\ 
The detailed description of the method and the usefulness of discrete symmetries which 
are applicable to both DI and glue will be presented in Sec.~\ref{subsec:discretesymmetries}.

\subsection{Ratios of Correlation functions}
\label{subsec:ratioscorrelationfunctions}
%
%
After evaluating two-point and three-point correlation functions,\ we take the following 
ratios between three-point to two-point functions, which at a large time separation involve 
the combinations of $T_1(q^2),\ T_2(q^2)$ and $T_3(q^2)$
\eqarray{
  & &\frac{\mbox{Tr}\left[\Gamma^{\unpol,\pol} G^{N {\mathcal T}_{4i} N}
      (\vec{p}\,',t_2;\vec{q},t_1; \vec{p},t_0)\right]}
  {\mbox{Tr}\left[\Gamma^{\unpol} G^{NN}(\vec{p}\,',t_2; t_0)\right]} \times
  \sqrt{\frac{\mbox{Tr}\left[\Gamma^{\unpol}  G^{NN}(\vec{p},t_2-t_1+t_0; t_0)\right]}
    {\mbox{Tr}\left[\Gamma^{\unpol} G^{NN}(\vec{p}\,',t_2-t_1+t_0; t_0)\right]}} 
  \nonumber\\
  &\times& \sqrt{\frac{\mbox{Tr}\left[\Gamma^{\unpol} G^{NN}(\vec{p}\,',t_1; t_0)\right]}
    {\mbox{Tr}\left[\Gamma^{\unpol} G^{NN}(\vec{p},t_1; t_0)\right]}\times
    \frac{\mbox{Tr}\left[\Gamma^{\unpol} G^{NN}(\vec{p}\,',t_2; t_0)\right]}
    {\mbox{Tr}\left[\Gamma^{\unpol} G^{NN}(\vec{p},t_2; t_0)\right]}} \nonumber\\
  & & \nonumber\\
  & &
  {\overrightarrow{\hspace{\z}}}\hspace{-\z}\raisebox{2ex}
  {$(t_1 - t_0),\, (t_2 - t_1) \gg 1/\Delta E$} 
  \hspace{4mm}
  \frac{\left[a_1T_1(q^2) + a_2T_2(q^2) + a_3T_3(q^2)\right]}
  {4\sqrt{E_{p'} (E_{p'}+m)E_{p}(E_{p}+m)}} .
  \label{eq:three_pt_2_two_pt_rat_1}
}
Since $T_1^q (q^2)$ and $T_2^q (q^2)$ for the CI are shown to have quite different $q^2$ 
behavior~\cite{Hagler:2003jd,Gockeler:2003jfa,Bratt:2010jn},\ we need to separately 
extrapolate $T_1(q^2)$ and $T_2(q^2)$ to $q^2 \longrightarrow 0$ (we shall do this both 
for CI and DI as well as for the glue contribution).\ In order to achieve this,\ we shall combine 
results from different kinematics for both the polarized and unpolarized three-point functions 
into the ratios in Eq.~(\ref{eq:three_pt_2_two_pt_rat_1}) at a particular $q^2$.\ The ratios 
then appear as different combinations in different $a_i$'s from which one can separate 
$T_1(q^2),\ T_2(q^2)$ and $T_3(q^2)$,\ and then extrapolate $T_1(q^2)$ and $T_2(q^2)$ in 
$q^2$ to obtain $T_1(0)$ and $T_2(0)$.\ The procedure to extract $T_1(q^2),\ T_2(q^2)$ and 
$T_3(q^2)$ is discussed in detail in Sec.~\ref{sec:separation_of_t1_t2_t3}.
%
\subsubsection{Special Case:\ $\vec{p}\,' = 0$ or $\vec{p}  = 0$}
%
%
\noindent
If we consider the special case with $\vec{p}\,' = 0$ for the polarized three-point functions,\ 
we obtain
\eqarray{
  & & \frac{\mbox{Tr}\left[\Gamma_l^{\pol}  G^{N {\mathcal T}_{4i} N}
      (\vec{0},t_2;\vec{q},t_1; -\vec{q},t_0)\right]}
  {\mbox{Tr}\left[\Gamma^{\unpol}  G^{NN}(\vec{0},t_2;t_0)\right]} \cdot
  \frac{\mbox{Tr}\left[\Gamma^{\unpol}  G^{NN}(\vec{0},t_1;t_0)\right]}
  {\mbox{Tr}\left[\Gamma^{\unpol}  G^{NN}(\vec{q},t_1;t_0)\right]}\nonumber\\
  & &
  {\overrightarrow{\hspace{\z}}}\hspace{-\z}\raisebox{2ex}
  {$(t_1 - t_0),\, (t_2 - t_1) \gg 1/\Delta E$} \hspace{4mm}
  \frac{-i}{4}\, \epsilon_{ijl}\, q_j \left[T_1 + T_2\right] (q^2) ,
  \label{eq:three_pt_2_two_pt_sp1_rat_1}
}
where $\left[T_1 + T_2\right] (q^2) = T_1(q^2) + T_2(q^2)$.\ Similarly if we consider 
$\vec{p} = 0$,\ we get
\eqarray{
  & & \frac{\mbox{Tr}\left[\Gamma_l^{\pol}  G^{N {\mathcal T}_{4i} N}
      (\vec{q},t_2;\vec{q},t_1; \vec{0},t_0)\right]}
  {\mbox{Tr}\left[\Gamma^{\unpol}  G^{NN}(\vec{q},t_2;t_0)\right]} \cdot
  \frac{\mbox{Tr}\left[\Gamma^{\unpol}  G^{NN}(\vec{q},t_1;t_0)\right]}
  {\mbox{Tr}\left[\Gamma^{\unpol}  G^{NN}(\vec{0},t_1;t_0)\right]}\nonumber\\
  & &
  {\overrightarrow{\hspace{\z}}}\hspace{-\z}\raisebox{2ex}
  {$(t_1 - t_0),\, (t_2 - t_1) \gg 1/\Delta E$} \hspace{4mm}
  \frac{-i}{4}\, \frac{E_{p'} + m}{m}\, \epsilon_{ijl}\, q_j \left[T_1 + T_2\right] (q^2) .
  \label{eq:three_pt_2_two_pt_sp2_rat_1}
}
In the unpolarized case,\ the three-point functions vanish when either $\vec{p}\,' = 0$ or 
$\vec{p} = 0$.

\noindent
We shall use the 
Eqs.~(\ref{eq:three_pt_2_two_pt_sp1_rat_1})~and~(\ref{eq:three_pt_2_two_pt_sp2_rat_1})
to check the extracted values of $T_1 (q^2)$ and $T_2 (q^2)$ by comparing 
$\left[ T_1 (q^2) + T_2 (q^2) \right]$ against $\left[T_1 + T_2\right](q^2)$ obtained 
directly at comparable $q^2$.
%
%
\subsubsection{Special Case:\ $\vec{p}\,' = \vec{p}$}
\label{subsec:ratioscorrelationfunctions_p.eq.p'}
%
%
If we consider the forward matrix element where $\vec{p}\,' = \vec{p}$ and take  the 
following ratio with unpolarized three-point functions,\ we obtain
\eqarray{
  \frac{\mbox{Tr}\left[\Gamma^{\unpol} G^{N {\mathcal T}_{4i} N}
      (\vec{p}\,',t_2;\vec{0},t_1; \vec{p}\,',t_0)\right]}
  {\mbox{Tr}\left[\Gamma^{\unpol} G^{NN}(\vec{p}\,',t_2)\right]}\hspace{1mm}
  &
  {\overrightarrow{\hspace{\z}}}\hspace{-\z}\raisebox{2ex}
  {$(t_1 - t_0),\, (t_2 - t_1) \gg 1/\Delta E$}
  &\hspace{4mm} T_1 (0)\, 
  =\, \langle x \rangle .
  \label{eq:first_mom_rat_1}
}
For the polarized case,\ the above ratio vanishes.\ We see that Eq.~(\ref{eq:first_mom_rat_1}) 
gives us the first moment of the parton distribution,\ or $T_1 (0)$,\ directly which has been 
calculated on the same lattice in~\cite{Deka:2008xr}.\ The value of $T_1 (0)$ obtained from 
Eq.~(\ref{eq:first_mom_rat_1}) can be checked against the $q^2$ extrapolated value of $T_1 (0)$ 
obtained from Eq.~(\ref{eq:three_pt_2_two_pt_rat_1}) or vice versa.

\smallskip

\noindent
Since Eq.~(\ref{eq:first_mom_rat_1}) allows a direct determination of $T_1 (0)$ without 
requiring one to perform $q^2 \rightarrow 0$ extrapolation,\ this results in a much clearer 
signal for $T_1 (0)$ as compared to that obtained from Eq.~(\ref{eq:three_pt_2_two_pt_rat_1}).\ 
The results from both the methods are presented in Sec.~\ref{sec:results} and,\ in fact,\ 
we shall combine the former one with the extrapolated value of $T_2(0)$ from 
Eq.~(\ref{eq:three_pt_2_two_pt_rat_1}) in order to construct $J^{q,g}$.\ Please note that 
these two methods for determining $T_1(0)$ are independent of each other since,\ in the 
$q^2 \rightarrow 0$ extrapolation method,\ we do not take into account the $q^2 = 0$ 
data point that comes from Eq.~(\ref{eq:first_mom_rat_1}).\ As mentioned earlier,\ this 
provides us a check for the value of $T_1(0)$.
%
%
%
\subsubsection{Ratios for Disconnected Insertions}
%
%
As mentioned in Sec.~\ref{subsec:threepoint_funcs},\ the sink time is fixed for CI.\ But in DI,\ 
the sink time need not be fixed,\ and we can sum over the insertion time,\ $t_1$,\ between 
the source and the sink time,\ i.e.\ from $t_1 = t_0 +1$ to $t_2 -1$ to gain more 
statistics~\cite{Dong:1995rx,Deka:2008xr,Maiani:1987by,Doi:2009sq}.\ Moreover,\ such 
summation helps in suppressing the excited state contamination~\cite{Deka:2008xr,
Maiani:1987by}.\ Similarly for gluons.\ Then the corresponding ratios at large time separation 
for Eqs.~(\ref{eq:three_pt_2_two_pt_rat_1}),\ (\ref{eq:three_pt_2_two_pt_sp1_rat_1}),\ 
(\ref{eq:three_pt_2_two_pt_sp2_rat_1}) and (\ref{eq:first_mom_rat_1}) become
\eqarray{
\label{eq:three_pt_2_two_pt_rat_2}
  & & \displaystyle\sum_{t_1 = t_0 +1}^{t_2 -1}
  \frac{\mbox{Tr}\left[\Gamma^{\unpol,\pol}  G^{N {\mathcal T}_{4i} N}
      (\vec{p}\,',t_2;\vec{q},t_1; \vec{p},t_0)\right]}
  {\mbox{Tr}\left[\Gamma^{\unpol}  G^{NN}(\vec{p}\,',t_2; t_0)\right]} \times
  \sqrt{\frac{\mbox{Tr}\left[\Gamma^{\unpol}  G^{NN}(\vec{p},t_2-t_1+t_0; t_0)\right]}
    {\mbox{Tr}\left[\Gamma^{\unpol}  G^{NN}(\vec{p}\,',t_2-t_1+t_0; t_0)\right]}} \nonumber \\
  &\times& \sqrt{\frac{\mbox{Tr}\left[\Gamma^{\unpol}  G^{NN}(\vec{p}\,',t_1; t_0)\right]}
    {\mbox{Tr}\left[\Gamma^{\unpol}  G^{NN}(\vec{p},t_1; t_0)\right]}\cdot
    \frac{\mbox{Tr}\left[\Gamma^{\unpol}  G^{NN}(\vec{p}\,',t_2; t_0)\right]}
         {\mbox{Tr}\left[\Gamma^{\unpol}  G^{NN}(\vec{p},t_2; t_0)\right]}} \nonumber\\
  & & \nonumber\\
  & &
  {\overrightarrow{\hspace{\z}}}\hspace{-\z}\raisebox{2ex}
  {$(t_1 - t_0),\, (t_2 - t_1) \gg 1/\Delta E$} 
  \hspace{2mm}
   \frac{\left[a_1T_1(q^2) + a_2T_2(q^2) + a_3T_3(q^2)\right]}
        {4 \sqrt{E_{p'} (E_{p'} + m) E_p (E_p + m)}} \times t_2 + {\rm const.}\, ,\\
  & & \nonumber\\
  & & \nonumber\\
\label{eq:three_pt_2_two_pt_sp1_rat_2}
  & & \displaystyle\sum_{t_1 = t_0 +1}^{t_2 -1}
  \frac{\mbox{Tr}\left[\Gamma_l^{\pol}  G^{N {\mathcal T}_{4i} N}
  (\vec{0},t_2;\vec{q},t_1; -\vec{q},t_0)\right]}
  {\mbox{Tr}\left[\Gamma^{\unpol}  G^{NN}(\vec{0},t_2;t_0)\right]} \cdot
  \frac{\mbox{Tr}\left[\Gamma^{\unpol}  G^{NN}(\vec{0},t_1;t_0)\right]}
  {\mbox{Tr}\left[\Gamma^{\unpol}  G^{NN}(\vec{q},t_1;t_0)\right]}\nonumber\\
  & & \nonumber\\
  & &
  {\overrightarrow{\hspace{\z}}}\hspace{-\z}\raisebox{2ex}
  {$(t_1 - t_0),\, (t_2 - t_1) \gg 1/\Delta E$} \hspace{2mm}
  \frac{-i}{4}\, \epsilon_{ijl} q_j \left[T_1 + T_2\right] (q^2) \times t_2 + {\rm const.}\, ,\\
  & & \nonumber\\
  \label{eq:three_pt_2_two_pt_sp2_rat_2}
  & & \displaystyle\sum_{t_1 = t_0 +1}^{t_2 -1}
  \frac{\mbox{Tr}\left[\Gamma_l^{\pol}  G^{N {\mathcal T}_{4i} N}
      (\vec{q},t_2;\vec{q},t_1; \vec{0},t_0)\right]}
  {\mbox{Tr}\left[\Gamma^{\unpol}  G^{NN}(\vec{q},t_2;t_0)\right]} \cdot
  \frac{\mbox{Tr}\left[\Gamma^{\unpol}  G^{NN}(\vec{q},t_1;t_0)\right]}
 {\mbox{Tr}\left[\Gamma^{\unpol}  G^{NN}(\vec{0},t_1;t_0)\right]}\nonumber\\
 & &
 {\overrightarrow{\hspace{\z}}}\hspace{-\z}\raisebox{2ex}
 {$(t_1 - t_0),\, (t_2 - t_1) \gg 1/\Delta E$} \hspace{2mm}
 \frac{-i}{4}\, \frac{E_{p'} + m}{m}\, \epsilon_{ijl} q_j 
 \left[T_1 + T_2\right] (q^2) \times t_2 + {\rm const.}\, , \\
 & & \nonumber\\
 \label{eq:first_mom_rat_2}
 && \displaystyle\sum_{t_1 = t_0 +1}^{t_2 -1}
 \frac{\mbox{Tr}\left[\Gamma^{\unpol} G^{N {\mathcal T}_{4i} N}
     (\vec{p}\,', t_2; \vec{0},t_1; \vec{p}\,', t_0)\right]}
 {\mbox{Tr}\left[\Gamma^{\unpol} G^{NN}(\vec{p}\,',t_2)\right]}\hspace{1mm}
 {\overrightarrow{\hspace{\z}}}\hspace{-\z}\raisebox{2ex}
 {$(t_1 - t_0),\, (t_2 - t_1) \gg 1/\Delta E$}
 \hspace{2mm} \langle x \rangle  \times t_2 + {\rm const.}\nonumber\\
}
We then extract the slopes in $t_2$ and obtain $T_1(q^2), T_2(q^2)$,  
$\left[T_1 + T_2\right](q^2)_{q,g}$ and $\langle x \rangle_{q,g}$ in the DI the same way
as is done for the CI.

\section{Stochastic Estimator and Variance Reduction }
\label{sec:disconnected_insertions}

\subsection{Noise Estimate of Current Loop in DI, Gauge Field Tensor 
and Unbiased Subtraction}
\label{sub:di_loop}

As we mentioned in Sec.~\ref{subsec:threepoint_funcs},\ we adopt the complex $Z_2$ (or 
$Z_4$) noise~\cite{Dong:1993pk} to compute the current loop in DI,\ because $Z_N$ noise 
has been shown to have the minimum variance~\cite{bmt93,Dong:1993pk}.

\

As we can see from Eq.~(\ref{Glue_Tensor}),\ the calculation of the gauge field tensor 
involves trace over spin indices of the massless overlap Dirac operator~\cite{Liu:2007hq,
  Doi:2008hp}.\ Moreover,\ ${\mathcal T}_{\{4i\}g}$ involves trace over color indices (see 
Eq.~(\ref{g_contrib_def_1})),\ and the corresponding three-point function involves a sum 
over space.\ This is basically the same as the quark loop calculation. Since we adopt the 
Zolotarev approximation for the sign function in the overlap operator,\ it entails an inversion 
of the Wilson fermion kernel with multi-shifts~\cite{Chen:2003im}.\ Thus,\ we again use the
complex $Z_2$ noise to estimate the trace of Eq.~(\ref{Glue_Tensor}) to construct the
glue energy-momentum tensor in Eq.~(\ref{eq:g_contrib_def_2}).

\

It has been shown that the off-diagonal matrix element contributions to the variance can be 
reduced by subtracting a judiciously chosen set of traceless $N\times N$ matrices 
$Q^{(p)}$~\cite{Thron:1997iy},\ which satisfy 
$\displaystyle\sum^N_{n=1}Q^{(p)}_{n, n} =0,\ p=1,\cdots, P$.\ Then the expectation value is 
unchanged when $M^{-1}$ is substituted with 
$M^{-1} - \displaystyle\sum^P_{p=1}\lambda_p\, Q^{(p)}$ ($\lambda_p$ is a constant),\ and 
yet the variance can be reduced. A natural choice for the set of traceless matrices is the 
hopping parameter expansion of the inverse of the Wilson fermion matrix,\ 
$D_W$~\cite{Thron:1997iy},\ and it has been applied to the study of the quark orbital angular 
momentum~\cite{Mathur:1999uf},\ the flavor-singlet scalar meson~\cite{McNeile:2000xx},\ 
determinant estimate~\cite{Alexandru:2007bb},\ the quark momentum fraction 
$\langle x \rangle$~\cite{Deka:2008xr} and the strangeness electromagnetic form 
factor~\cite{Doi:2009sq}.\ We see a reduction of the errors by more than a factor of two with 
negligible cost. We shall adopt this unbiased subtraction with hopping expansion of the Wilson 
Dirac fermion to the fourth order.

\subsection{Discrete Symmetries and Transformations}
\label{subsec:discretesymmetries}

Since both the DI and glue operators are stochastically estimated,\ the signals for the 
corresponding three-point functions are usually noisy.\ In order to improve the signals,\ we 
take advantage of discrete symmetries to further reduce the variance from the gauge noise.
We will tap parity,\ $\gamma_5$ hermiticity,\ and charge-$\gamma_5$ hermiticity 
($CH$ transformation)~\cite{ber89,Draper:1988bp,Deka:2008xr} to filter out the 
noise contributions which would be zero with infinite statistics. This is the same idea as
the unbiased subtraction in Sec.~\ref{sub:di_loop}.

\subsubsection{Two-point Functions and Current Loop}
\label{subsubsec:discrete_sym_2pt_function_and_loop}

Since the three-point functions for DI are constructed by multiplying (or, correlating) the 
nucleon propagator with the current loop on each gauge configuration,\ we can consider 
the parity, $CH$, and $\gamma_5$ transformation properties of each of them.\ 
In Table~\ref{tab:discrete_trans_nuc},\ we show the outcome of parity and $CH$ 
transformations on the polarized and unpolarized nucleon propagators.\ Here we use the 
shorthand notation:\ 
$f (\vec{p}, t; t_0; U) = \mbox{Tr}\,[\Gamma^{\unpol}\, G^{NN} (\vec{p}, t; t_0; U)]$ and 
$g (\vec{p}, t; t_0; U) = \mbox{Tr}\,[\Gamma^{\pol}\, G^{NN} (\vec{p}, t; t_0; U)]$.
\begin{table}[h]
  \centering
  \renewcommand{\arraystretch}{1}
  \begin{tabular}{c|c|c}
    \hline\hline
        {\bf Nucleon}
        &\multirow{2}{*}{\bf Parity}
        & {\boldmath $CH$} \\
        {\bf Propagators} 
        & 
        & {\bf Transformations} \\
        \hline
        &&\\
        $f (\vec{p}, t; t_0; U)$
        & $f (-\vec{p}, t; t_0; U^p)$ 
        & $\Big[f (-\vec{p}, t; t_0; U^*)\Big]^*$ \\
        \hline
        &&\\
        $f (\vec{p}, t; t_0; U) + f (-\vec{p}, t; t_0; U)$
        & Even
        & $\Big[f (\vec{p}, t; t_0; U^*) + f (-\vec{p}, t; t_0; U^*)\Big]^*$\\
        \hline
        &&\\
        $g (\vec{p}, t; t_0; U)$ 
        & $g (-\vec{p}, t; t_0; U^p)$ 
        & $ -\, \Big[g (-\vec{p}, t; t_0; U^*)\Big]^*$ \\
        \hline\hline
  \end{tabular}
  \caption{Table showing the outcome of the parity and $CH$ transformations 
    on unpolarized and polarized nucleon propagators with equal and opposite momenta.\ 
$U^p$ and $U^*$ denote the parity and $C$ transformed gauge links,\ respectively.}
  \label{tab:discrete_trans_nuc}
\end{table}

\

\noindent
Similarly,\ the outcome of the parity,\ $\gamma_5$ \ and $CH$ transformations for the loop 
of the energy-momentum tensor in Eq.~(\ref{eq:disconnected_4i_f_1}) are shown in 
Table~\ref{tab:discrete_trans_loop}.
\begin{table}[h]
\centering
\renewcommand{\arraystretch}{1.2}
\begin{tabular}{c|c|c|c}
\hline\hline
\multirow{2}{*}{\bf Loop} 
& \multirow{2}{*}{\bf Parity} 
& {\boldmath$\gamma_5$} 
& {\boldmath$CH$}\\
& 
&{\bf Hermiticity }
& {\bf Transformations}\\
\hline
&&&\\
$L [t_1, \vec{q};U]$ 
& $ - L [t_1, - \vec{q};U^p]$  
& $\displaystyle\ \frac{+ i}{8a}\,
\displaystyle\sum_{\vec{x}_1} e^{i\vec{q} \cdot (\vec{x}_1 - \vec{x}_{0})}\,
\mbox{Re} \bigg[ L [x_1;U] \bigg]$
& $- \bigg[L [t_1, - \vec{q};U^*]\bigg]^*$\\
\hline\hline
\end{tabular}
\caption{Table showing the outcome of the parity,\ $\gamma_5$
and $CH$ Transformations on the quark loop for the energy-momentum tensor
in Eq.~(\ref{eq:disconnected_4i_f_1}).}
\label{tab:discrete_trans_loop}
\end{table}

\subsubsection{Construction of Disconnected Three-point Functions}
\label{subsubsec:discrete_sym_di_3pt}

According to the CH theorem~\cite{bernard_1},\ after gauge averaging,\ the path integral 
for $\langle \mathcal{O}\rangle$  in QCD is either real or imaginary (except in the case with 
chemical potential).\ Using the transformation properties given in 
Tables~\ref{tab:discrete_trans_nuc}~and~\ref{tab:discrete_trans_loop},\ one can decide on 
the right combination of real and imaginary components of the nucleon propagator and the 
loop to satisfy the total parity and $CH$ transformation properties and the $\gamma_5$
hermiticity for the quark loop~\cite{Mathur:1999uf,Deka:2008xr,Doi:2009sq}.\ In this way,\ 
we obtain the unpolarized three-point functions (DI) as
\eqarray{
  &  & \mbox{Tr}\big{[}\Gamma^{\unpol}\, G^{{N {\mathcal T}_{4i} N}}
    (\vec{p}\,',t_2;\vec{q},t_1; \vec{p},t_0)\big{]}_{\rmsmall{DI}}\nonumber\\ 
  &=& \left(\frac{1}{8a}\right) \left\langle
    \bigg\{\displaystyle\sum_{\vec{x}_2} \cos(\vec{p}\,' \cdot (\vec{x}_2 - \vec{x}_0))\, 
    \mbox{Re}\Big[N^{\unpol}[x_2;U]\Big]
    \displaystyle\sum_{\vec{x}_1}\sin(\vec{q} \cdot (\vec{x}_1 - \vec{x}_{0}))\,
    \mbox{Re}\Big[L [x_1;U]\Big]\right.\nonumber\\
  & & \left. -\, \displaystyle\sum_{\vec{x}_2} \sin(\vec{p}\,' \cdot (\vec{x}_2 - \vec{x}_0))\, 
    \mbox{Re}\Big[N^{\unpol}[x_2;U]\Big]
    \displaystyle\sum_{\vec{x}_1}\cos(\vec{q} \cdot (\vec{x}_1 - \vec{x}_{0}))\,
    \mbox{Re}\Big[L [x_1;U]\Big]\bigg\}\right\rangle ,
  \label{eq:disconnected_3pt_unpol}
}
and the polarized three-point functions (DI) as
\eqarray{
  &  & \mbox{Tr}\big{[}\Gamma_l^{\pol}\, G^{{N {\mathcal T}_{4i} N}}
    (\vec{p}\,',t_2;\vec{q},t_1; \vec{p},t_0)\big{]}_{\rmsmall{DI}}\nonumber\\ 
  &=& \left(\frac{i}{8a}\right) \left\langle
    \bigg\{\displaystyle\sum_{\vec{x}_2} \cos(\vec{p}\,' \cdot (\vec{x}_2 - \vec{x}_0))\, 
    \mbox{Im}\Big[N_l^{\pol}[x_2;U]\Big]
    \displaystyle\sum_{\vec{x}_1}\sin(\vec{q} \cdot (\vec{x}_1 - \vec{x}_{0}))\,
    \mbox{Re}\Big[L [x_1;U]\Big] \right.\nonumber\\
    & &\left. -\, \displaystyle\sum_{\vec{x}_2} \sin(\vec{p}\,' \cdot (\vec{x}_2 - \vec{x}_0))\, 
    \mbox{Im}\Big[N_l^{\pol}[x_2;U]\Big]
    \displaystyle\sum_{\vec{x}_1}\cos(\vec{q} \cdot (\vec{x}_1 - \vec{x}_{0}))\,
    \mbox{Re}\Big[L [x_1;U]\Big]\bigg\}\right\rangle .
    \label{eq:disconnected_3pt_pol}
  }

\section{Choice of Momenta}
\label{sec:choice_of_mom}

The momenta we shall choose for computing the first moment of the momentum fraction 
carried by quarks for both CI and DI have been discussed in detail in~\cite{Deka:2008xr}.\ 
For the case of glue,\ we shall use the same momenta as in the case of DI.

\smallskip

For angular momenta,\ we have discussed earlier that (see 
Sec.~\ref{subsec:ratioscorrelationfunctions}) we need to combine several kinematics 
into the ratios in Eq.~(\ref{eq:three_pt_2_two_pt_rat_1}) for CI or in 
Eq.~(\ref{eq:three_pt_2_two_pt_rat_2}) for DI at a particular $q^2$ 
from which one can separate $T_1(q^2),\ T_2(q^2)$ and $T_3(q^2)$.\ For this purpose,\ we 
first take several momenta to set up the suitable kinematics.\ Since both the two-point and 
three-point functions are subject to larger noise with higher momenta,\ we have limited 
ourselves to momenta not exceeding $2$ (in lattice units).\ With these momenta under 
consideration,\ we can construct four different values of $q^2$ for which 
$\vec{p} \neq \vec{p}\,' \neq \vec{q} \neq 0$.\ Since the momentum projection is folded in 
the sequential source at the sink time $t_2$ in the CI computation,\ we have chosen only the 
cases for which $\vec{p}\,' = ( 1, 0, 0 )$ in order to reduce the computational cost.\ In 
contrast,\ the computation of the valence quark propagators in DI is separate from the loop 
computation in each configuration;\ this means that the momentum in the nucleon two-point 
functions can be chosen independently of the momentum transfer carried by the costly loop 
calculation,\ only constrained by momentum conservation.\ This allows us to choose all the 
available momenta at the same computational cost.\ Similar is the case for the glue 
contributions.

\section{Separation of {\boldmath $T_1$},\ {\boldmath $T_2$},\ 
  {\boldmath $T_3$}}
\label{sec:separation_of_t1_t2_t3}

In this section,\ we will discuss how to separate $T_1$,\ $T_2$ and $T_3$ at a particular value 
of $q^2$ (For details,\ see Appendix~\ref{appsec:kine_eqs}).\ Using the available momenta,\ 
we obtain several ratios of three-point to two-point functions (both polarized and unpolarized) 
for all the three directions of the operator,\ ${\mathcal T}_{4i}$,\ at every $q^2$.\ We then 
average over the ratios with the same coefficients,\ $a_i$'s,\ and extract them either by fitting 
a constant (for CI) or by fitting a slope (for DI and glue).\ This results in a fewer but more than 
three different equations which contain $T_1$,\ $T_2$ and $T_3$ with different coefficients 
$a_i$'s.\ Though these equations are analytically different,\ numerically they are correlated 
since they are computed on the same set of configurations.\ While solving for $T_1$,\ $T_2$ 
and $T_3$,\ such correlations must be taken into account.\ Therefore,\ we construct a 
covariance matrix,\ $C$,\ between these equations for every $q^2$.\ We then construct the 
following $\chi^2$ as
\eqarray{
  \chi^2 
  &=& \displaystyle\sum_{ij}^N 
  \Big[R_i - (a_{1,i}\, T_1 + a_{2,i}\, T_2 + a_{3,i} T_3)\Big]\, 
  C_{ij}^{-1}\, 
  \Big[R_j - (a_{1,j}\, T_1 + a_{2,j}\, T_2 + a_{3,j} T_3)\Big] ,
\label{eq:chi_sq_ratio}
}
where $N$ is the number of equations,\ and $R_i$'s are the fitted values of the ratios.\  
Minimizing the $\chi^2$ in Eq.~(\ref{eq:chi_sq_ratio}) w.r.t. $T_1$,\ $T_2$ and $T_3$,\ 
we obtain the following three equations
\eqarray{
  \left[
    \begin{array}{c}
      R'_1 \\
      R'_2 \\
      R'_3 
    \end{array}
    \right] 
  &=& \left[
    \begin{array}{ccc} 
      a^1_1 & a^1_2 & a^1_3 \\
      a^2_1 & a^2_2 & a^2_3 \\
      a^3_1 & a^3_2 & a^3_3 \\
    \end{array}
    \right]\,
  \left[
  \begin{array}{c}
    T_1 \\
    T_2 \\
    T_3 
  \end{array}
  \right] ,
\label{eq:3_ind_eq_ratio}
}
where
\eq{
a^m_k = 2\, a_{m,i}\, C_{ij}^{-1}\, a_{k,j},\,
R'_k = 2\, a_{k,i}\, C_{ij}^{-1}\, R_j, \hspace{15mm}
(m, k = 1, 2, 3),
\label{eq:def_a.r'}
}
and the sum over $i, j$ is implicitly implied.\ Solving the system of equations in 
Eq.~(\ref{eq:3_ind_eq_ratio}),\ we can separate $T_1$,\ $T_2$ and $T_3$ at that $q^2$.
%

\section{Numerical Parameters}
\label{sec:numerical_param}

We use 500 gauge configurations on a $16^3 \times 24$ lattice generated with Wilson action 
at $\beta=6.0$ in the quenched approximation.\ They are produced by the pseudo-heatbath 
algorithm with $10,000$ sweeps between consecutive configurations.\ The values of the 
hopping parameter we have used are $\kappa = 0.154$,\ $0.155$ and $0.1555$.\ The critical 
hopping parameter,\ $\kappa_c = 0.1568$ is obtained by a linear extrapolation to the zero 
pion mass~\cite{Dong:1995ec}.\ Using the nucleon mass to set the lattice spacing at 
$a = 0.11$~fm,\ the corresponding pion masses are $650(3)$,\ $538(4)$,\ and $478(4)$~MeV,\ 
and the nucleon masses are $1291(9)$,\ $1159(11)$, and $1093(13)$~MeV,\ respectively.\ In 
the present work,\ we use periodic boundary condition in the spatial directions.\ In the temporal 
direction,\ Dirichlet boundary condition is imposed at $t = 1$ and $t = 24$.\ This provides a 
larger time separations than those available with periodic boundary conditions.

\medskip

\noindent
The quark loops for DI and overlap operator for glue are computed separately using complex 
$Z_2$ noise vector~\cite{Dong:1993pk}.\ The number of noise vectors we use for DI is $500$ 
on each gauge configuration.\ Also for the case of quarks,\ we shall define two $\kappa$'s for 
the quark mass:\ $\kappa_v$ for valence quarks,\ and $\kappa_{\rmsmall{loop}}$ for quarks in 
the current loop in the case of DI.\ For the strange quark currents,\ we have fixed 
$\kappa_{\rmsmall{loop}} = 0.154$ which is close to the strange quark mass as determined 
from the $\phi$~meson mass,\ and $\kappa_v$ takes the values $0.154$,\ $0.155$ and 
$0.1555$.\ For up and down quarks,\ we consider equal masses for valence quarks and 
quarks in the current loop,\ i.e.\ $\kappa_{\rmsmall{loop}} = \kappa_v = 0.154, 0.155$, and 
$0.1555$.\ The source time for the quark propagators is fixed at $t_0 = 4$.\ In the case of CI,\ 
the sink time is fixed at $t_2 = 16$.

\medskip

\noindent
We estimate the gauge field tensor from the overlap operator stochastically with two complex 
$Z_2$ noise vectors on each configuration,\ but with dilution in color and spin indices.\ For the 
space-time points,\ we perform a dilution with multiple grids to cover the whole space-time 
points. The points on the grid are separated by two sites on top of odd/even dilution.\ 
Therefore,\ the “taxi-driver distance” equals 4 in our case.\ The reason behind the grid dilution 
approach is that,\ unlike the quark loop,\ the overlap operator is exponentially local with a range 
of fall-off to be about two lattice spacing in the taxi driver distance.

\medskip

\noindent
We use multiple nucleon sources ($16$ in this work) to increase the statistics in the cases 
of DI and glue.\ We correlate all the corresponding two-point functions with the already 
computed DI and the glue energy-momentum tensor.\ This has shown to reduce the error 
significantly~\cite{Deka:2008xr,Doi:2009sq}.\ In the case of CI,\ we use only one nucleon 
source.

\medskip

\noindent
The error analysis is performed by using the jackknife procedure.\ The correlations among 
different quantities are taken into account by constructing the corresponding covariance 
matrices.\ In order to extract various physical quantities,\ we use correlated least-$\chi^2$ 
fits.\ To determine $T_1 (0)$ and $T_2 (0)$,\ we first separate $T_1 (q^2)$,\ $T_2 (q^2)$ and 
$T_3 (q^2)$ at finite $q^2$ using the method discussed in Sec.~\ref{sec:separation_of_t1_t2_t3} 
for every jackknife sample.\ $T_1 (0)$ and $T_2 (0)$ are then obtained by extrapolating 
$q^2$ to zero with a dipole form.\ Alternatively,\ $T_1(0)$ can be directly computed from 
the forward matrix element as discussed in Sec.~\ref{subsec:ratioscorrelationfunctions_p.eq.p'}.\ 
We should point out that we do not take into account the forward matrix value of $T_1(0)$ 
when we perform the $q^2 \rightarrow 0$ extrapolation for $T_1 (q^2)$.\ The values of $T_1(0)$ 
obtained from both the methods are consistent within errors and presented in 
Sec.~\ref{sec:results}.\ Since $T_1 (0)$ obtained from forward matrix element is more 
precise with a smaller error,\ we shall use it in the following discussion as well as combining 
with $q^2$-extrapolated value of $T_2 (0)$ to construct $2 J$.

\section{Results and Discussion}
\label{sec:results}

\subsection{Connected Insertions}
\label{subsec:results_ci}
\begin{figure}[h]
\centering
\subfigure[]
{\rotatebox{270}{\includegraphics[width=0.34\hsize]{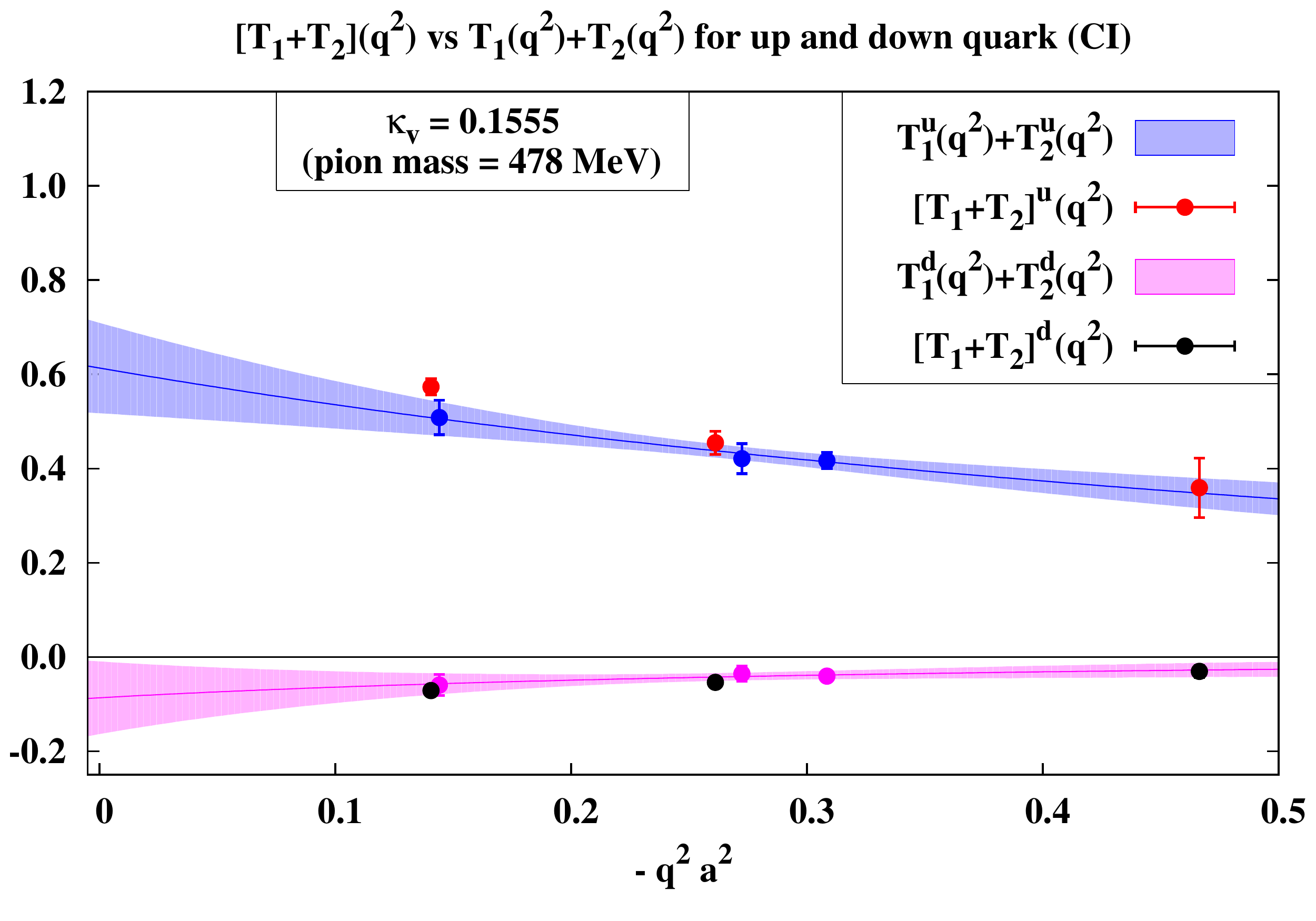}}
\label{fig:quark_ang_mom_CT_1}}
\subfigure[]
{\rotatebox{270}{\includegraphics[width=0.34\hsize]{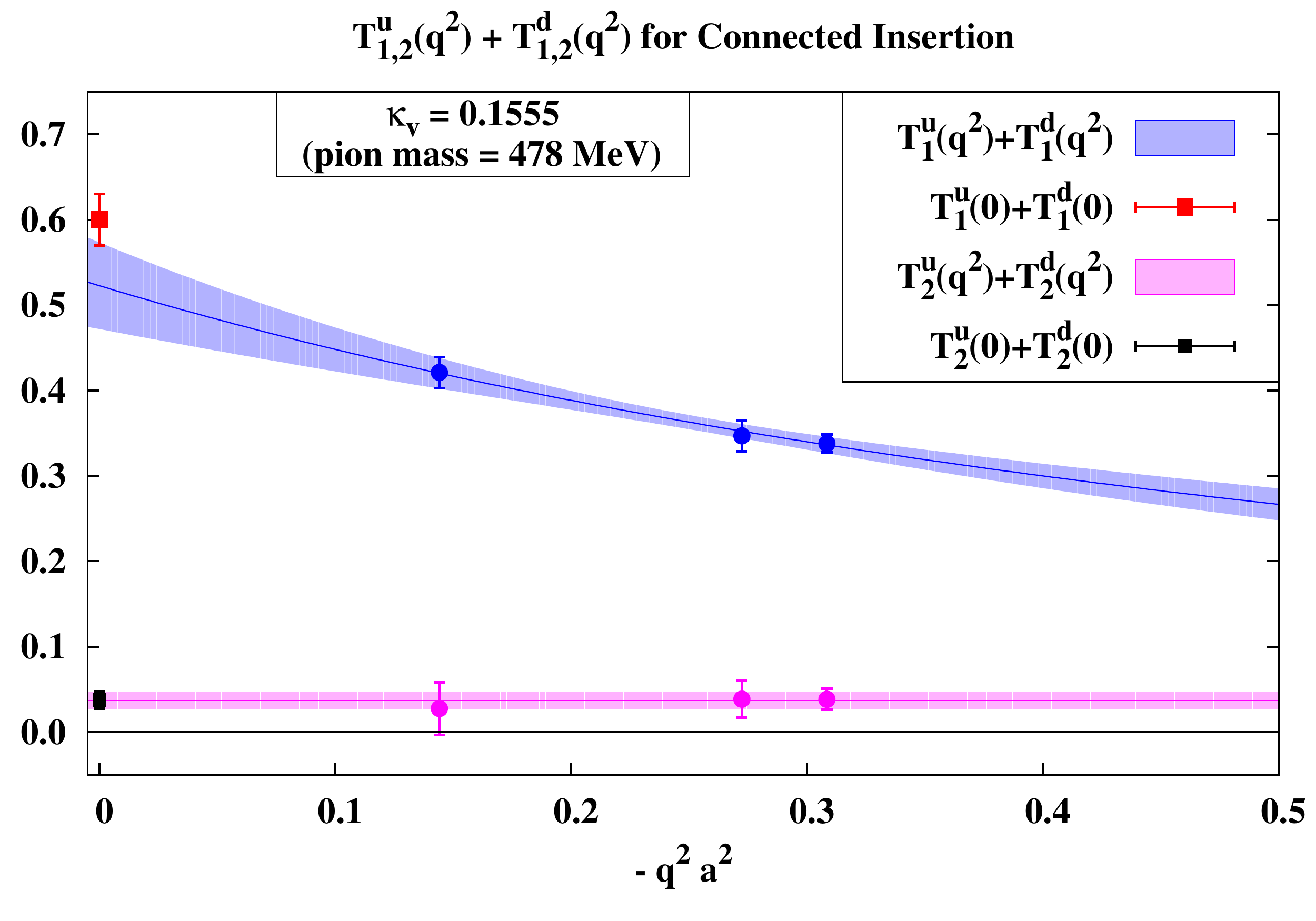}}
 \label{fig:quark_ang_mom_CT_2}}
\caption{CI plots at $\kappa = 0.1555$.\ 
  (a) The sum of $T_1(q^2)$ and $T_2(q^2)$,\ extracted from 
   Eqs.~(\ref{eq:three_pt_2_two_pt_rat_1}) and (\ref{eq:3_ind_eq_ratio}) along 
   with error bands from the dipole fit,\ is compared to $\left[T_1 + T_2\right](q^2)$ obtained 
   from Eqs.~(\ref{eq:three_pt_2_two_pt_sp1_rat_1}) and 
   (\ref{eq:three_pt_2_two_pt_sp2_rat_1}) at comparable $q^2$ values for $u$ and $d$ quarks 
   in the CI.\
   (b) The sum of $u$ and $d$ quark contributions for $T_1(q^2)$ and $T_2(q^2)$.\ The 
   red square at $q^2 = 0$ is $\left[T_1^u (0) + T_1^d (0)\right]$ which is obtained from 
   forward matrix elements using Eq.~(\ref{eq:first_mom_rat_1}).\ The black square at 
   $q^2 = 0$ is $\left[T_2^u (0) + T_2^d (0)\right]$ which is obtained from dipole fit.\ To 
   construct $J^{u+d}$~(CI),\ we add the values represented by the red and black squares.
}
\end{figure}

%
We first present our results for the CI.\ The analyses are straightforward extension of those 
in~\cite{Deka:2008xr}.\ In Fig.~\ref{fig:quark_ang_mom_CT_1},\ we plot
$\left[T_1^u (q^2) + T_2^u (q^2)\right]$ and $\left[ T_1^d(q^2) + T_2^d(q^2)\right]$ as 
functions of $q^2$ for $\kappa = 0.1555$, the smallest quark mass, where $T_1(q^2)$ and 
$T_2(q^2)$  are obtained by using Eqs.~(\ref{eq:three_pt_2_two_pt_rat_1}) and 
(\ref{eq:3_ind_eq_ratio}).\ We also plot $\left[T_1 + T_2\right]^u (q^2)$ and 
$\left[T_1 + T_2\right]^d (q^2)$ obtained directly from 
Eqs.~(\ref{eq:three_pt_2_two_pt_sp1_rat_1}) and (\ref{eq:three_pt_2_two_pt_sp2_rat_1}) 
at slightly different but comparable $q^2$\,'s.\ 
We see that the latter agrees within $2\sigma$ 
of the error band of the former which is obtained from a dipole fit in~$q^2$.\ This is a cross 
check of our procedure of extracting $T_1(q^2)$ and $T_2(q^2)$.\ We also show 
$\left[T_1^u (q^2) + T_1^d (q^2)\right]$ and$\left[T_2^u(q^2) + T_2^d(q^2)\right]$ and 
their error bands  in Fig.~\ref{fig:quark_ang_mom_CT_2}.\ Also plotted is 
$\left[T_1^u(0) + T_1^d(0)\right]$ from Eq.~(\ref{eq:first_mom_rat_1}).\ We see that 
its error is smaller than that from the separately extrapolated $T_1^u(0)$ and $T_1^d(0)$.\ 
Thus we shall use $\left[T_1^u(0) + T_1^d(0)\right]$ obtained from 
Eq.~(\ref{eq:first_mom_rat_1}) and combine with $\left[T_2^u(0) + T_2^d(0)\right]$ 
obtained from the dipole fit to get the angular momentum $J^q$ for the CI.\ We follow similar 
procedure for other $\kappa_v$ values.

\subsection{Disconnected Insertions}
\label{subsec:results_di}
%
%
\begin{figure}[h]
  \centering
\subfigure[]
{\rotatebox{270}{\includegraphics[width=0.33\hsize]{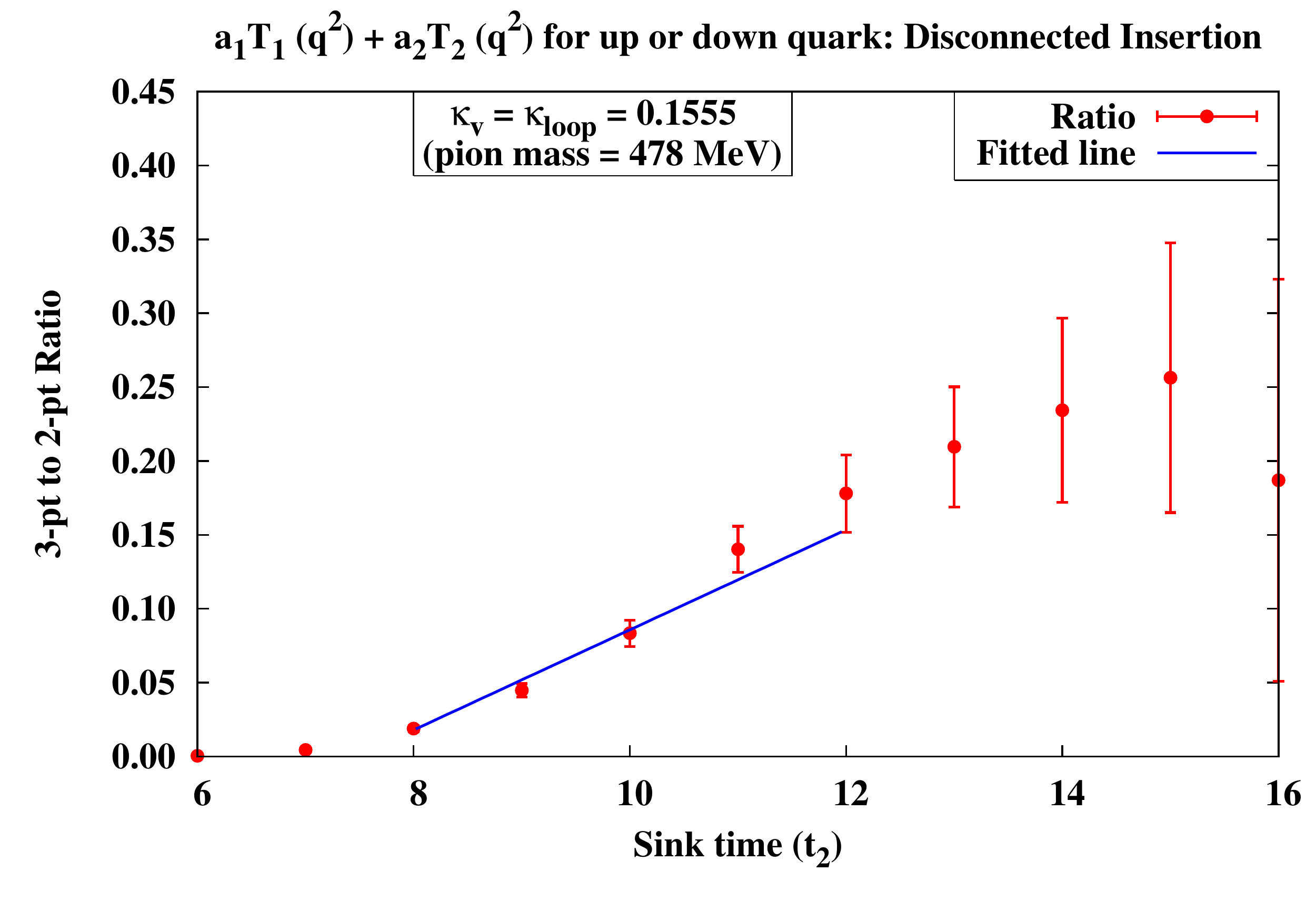}}
  \label{fig:slope_t1t2t3_1555_DI}}
\hspace{2mm}
\subfigure[]
{\rotatebox{270}{\includegraphics[width=0.33\hsize]{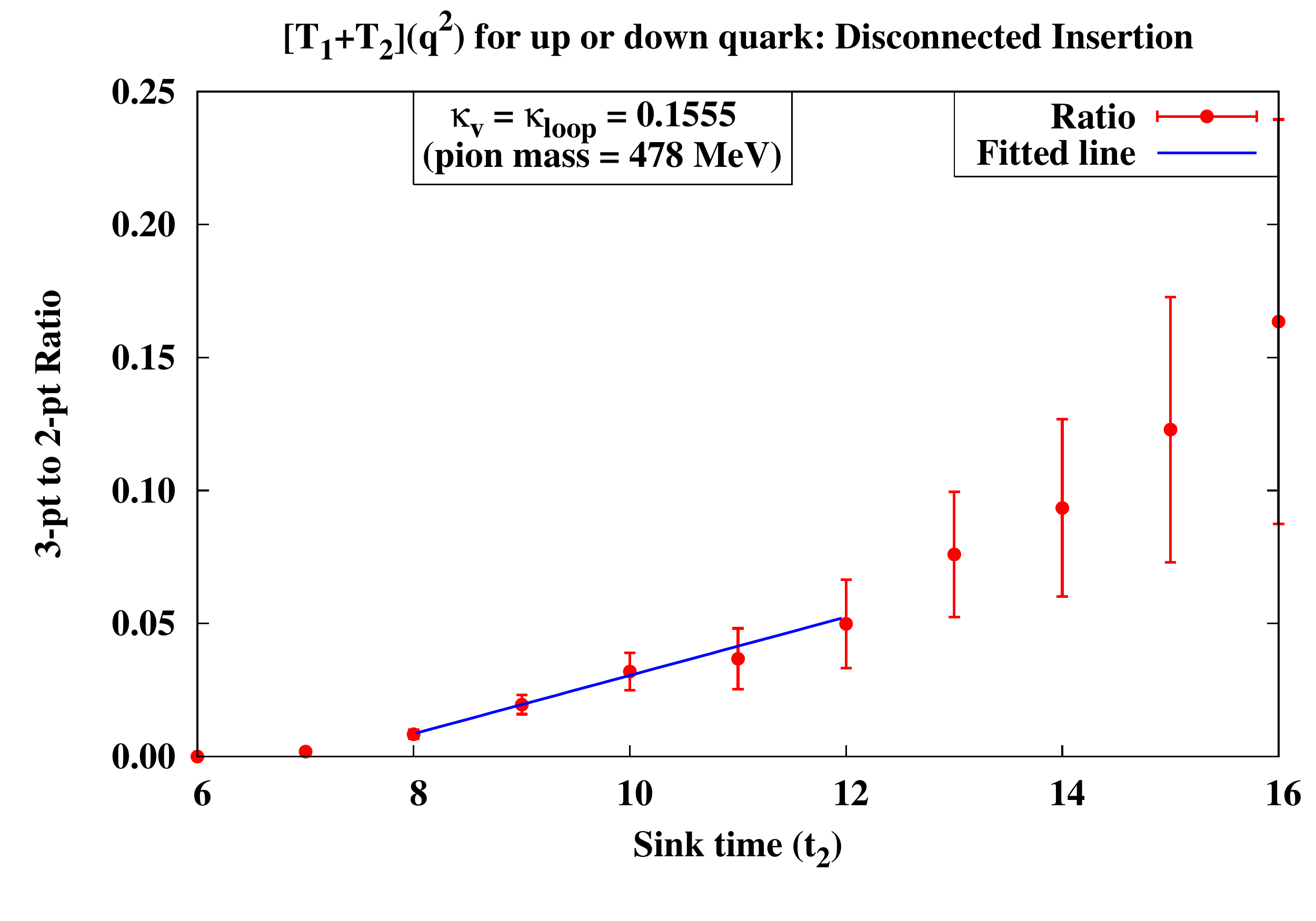}}
  \label{fig:slope_1555_DI}}
\subfigure[]
{\rotatebox{270}{\includegraphics[width=0.33\hsize]{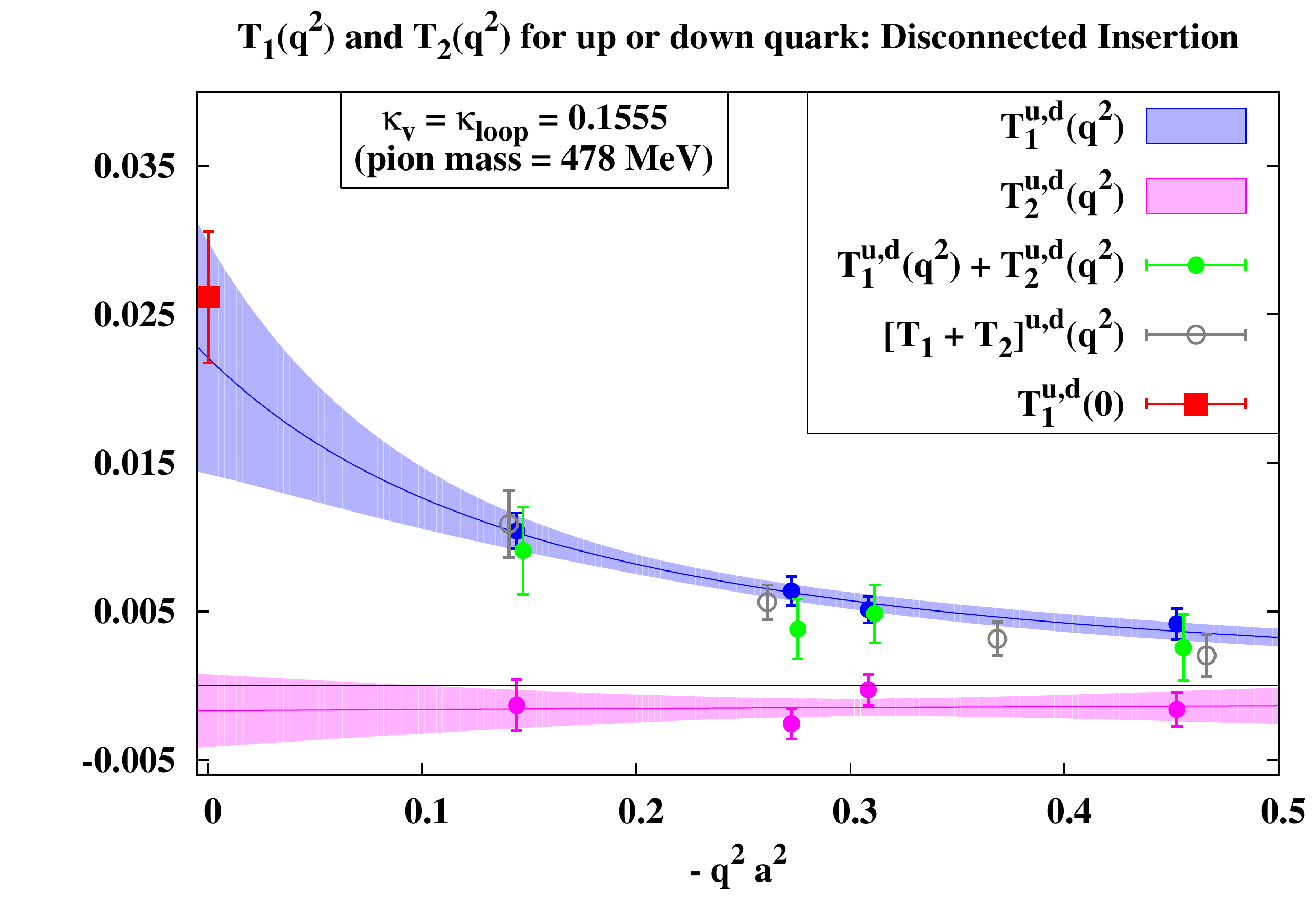}}
  \label{fig:T1T2_DI}}
\hspace{2mm}
\subfigure[]
{\rotatebox{270}{\includegraphics[width=0.33\hsize]{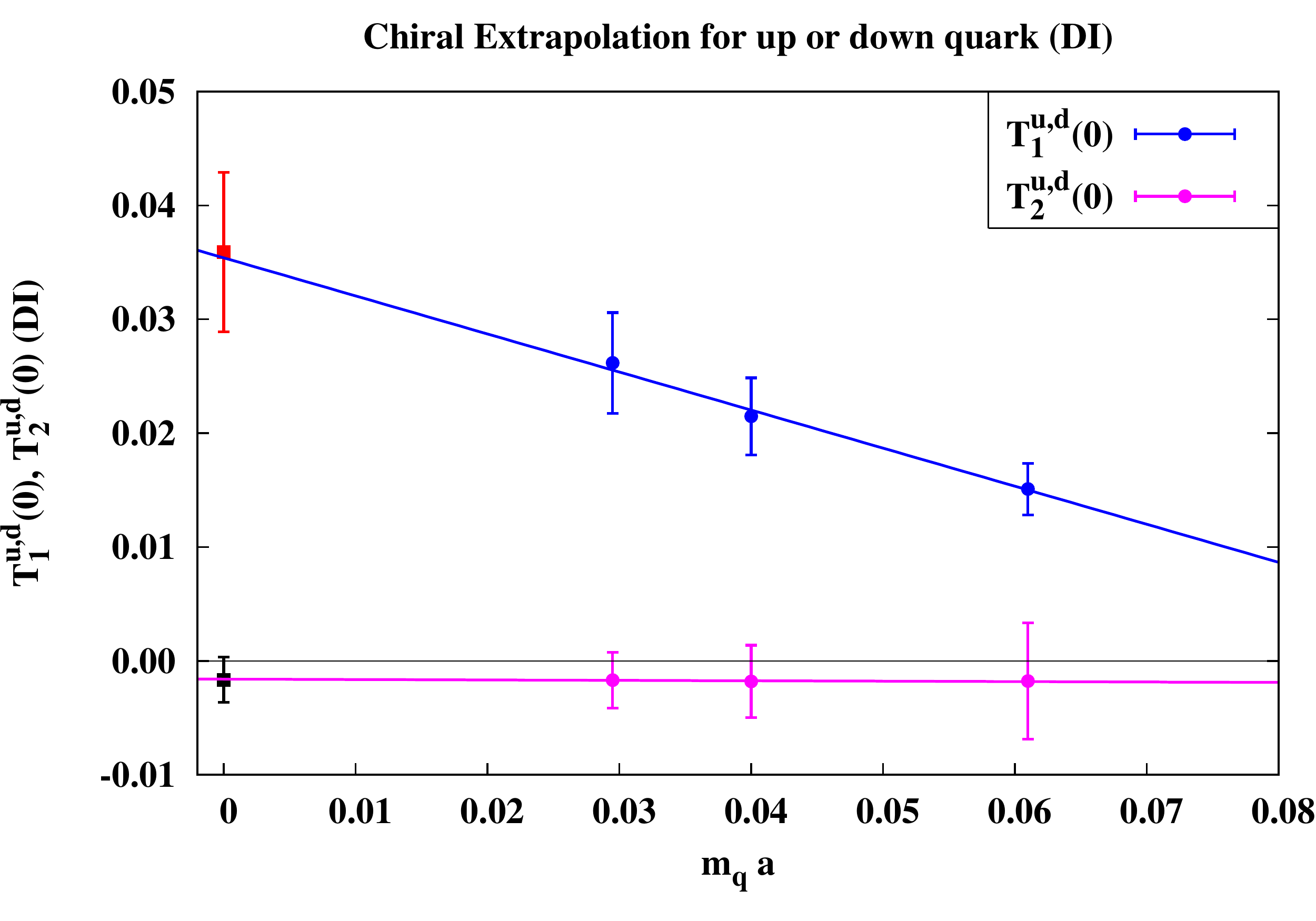}}
  \label{fig:chiral_fit_quark_ang_mom_DI}}
\caption{DI plots for $u, d$ at $\kappa_v = \kappa_{\rmsmall{loop}} = 0.1555$.\ 
  (a) One of the ratios in Eq.~(\ref{eq:three_pt_2_two_pt_rat_2}) plotted against the sink 
  time,\ $t_2$.\ The term with form factor $T_3 (q^2)$ does not appear in this particular 
  ratio.\ The slope is fitted to obtain $\left[a_1 T_1 (q^2) + a_2 T_2 (q^2)\right]^{u,d}$.\ 
  (b) The ratio in Eq.~(\ref{eq:three_pt_2_two_pt_sp1_rat_2}) plotted against the sink 
  time,\ $t_2$.\ The slope is fitted to obtain $\left[T_1 + T_2\right]^{u,d} (q^2)$.\ 
  (c) The sum of separately extracted $T_1(q^2)$ and $T_2(q^2)$ is compared with 
  $\left[T_1+ T_2\right](q^2)$.\ $T_1(0)$ (red square) is from the forward matrix element.\ 
  In order to construct $J$,\ the value represented by the red-square is used as $T_1(0)$.\ 
  (d) Chiral extrapolation of $T_1(0)$  and $T_2(0)$ for the $u/d$ quark.\ The red and black 
  squares in this figure represent chirally extrapolated values of $T_1(0)$  and $T_2(0)$,\ 
  respectively.\ Please note that they are not renormalized in this figure.}
\label{fig:DI}
\end{figure}

For the DI,\ we show one of the ratios in Eq.~(\ref{eq:three_pt_2_two_pt_rat_2}) plotted
against the sink time,\ $t_2$,\ in Fig.~\ref{fig:slope_t1t2t3_1555_DI} and the ratio in 
Eq.~(\ref{eq:three_pt_2_two_pt_sp1_rat_2}) similarly plotted 
in Fig.~\ref{fig:slope_1555_DI} with $\kappa_v = \kappa_{\rmsmall{loop}} = 0.1555$ at 
$q^2 = 0.144$.

We fit the slope from $t_2 =8$ where the two-point function begins to be dominated by the 
nucleon to $t_2 = 12$.\ We plot $\left[T_1 + T_2\right](q^2)$ so obtained in 
Fig.~\ref{fig:T1T2_DI},\ and compare them to $T_1(q^2) + T_2(q^2)$ extracted from $6$ 
combinations of $a_1 T_1(q^2) + a_2 T_2(q^2) + a_3 T_3 (q^2)$.\ We see that they are 
consistent with each other within errors.\ The error bands are from the dipole fits of 
$T_1(q^2)$ and $T_2(q^2)$.\ $T_1(0)$ (in red square) is from the forward matrix element 
which has smaller error than the $q^2$ extrapolated value of $T_1(0)$.\ Thus in a similar 
manner as in CI,\ we shall combine it with the extrapolated $T_2(0)$ to obtain the angular 
momentum $J^q$ (DI).\ We follow similar procedure for other $\kappa$ values,\ and strange 
quarks.

\

\noindent
Finally,\ we perform a linear chiral extrapolation of $\kappa_v$ to obtain $T_1(0) + T_2(0)$ 
for the $u, d$ quarks at the chiral limit.\ This is shown in 
Fig.~\ref{fig:chiral_fit_quark_ang_mom_DI}.\ For the strange quark,\ on the other hand,\ 
we fix the loop at $\kappa_{\rmsmall{loop}} = 0.154$,\ and then extrapolate the $\kappa_v$ to 
the chiral limit. 
%

\subsection{Glue}
\label{subsec:results_glue}

%
%
\begin{figure}[htbp]
\centering
\subfigure[]
{\rotatebox{270}{\includegraphics[width=0.335\hsize]
    {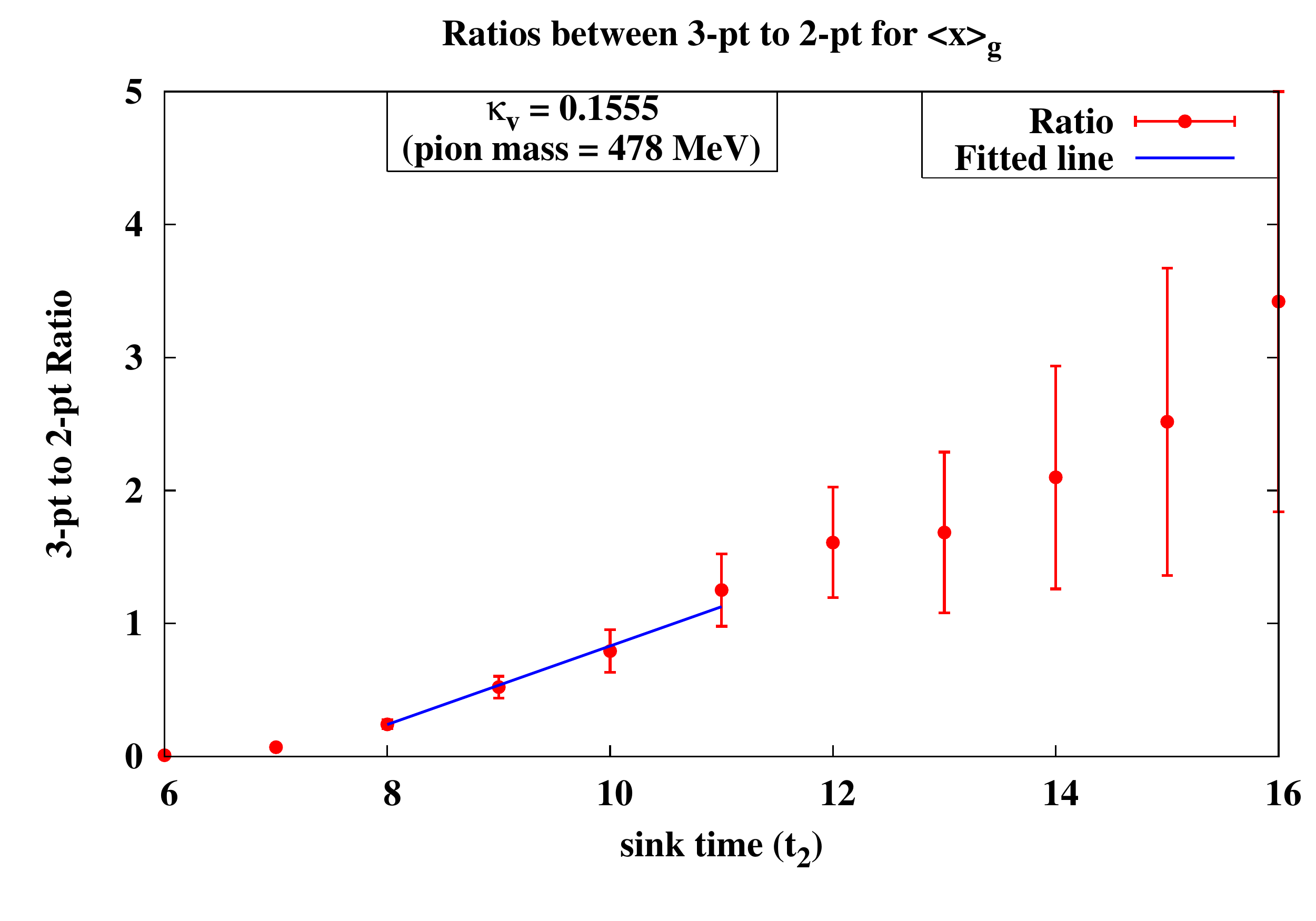}}
  \label{fig:slope_x_1555_g}}
\subfigure[]
{\rotatebox{270}{\includegraphics[width=0.335\hsize]
    {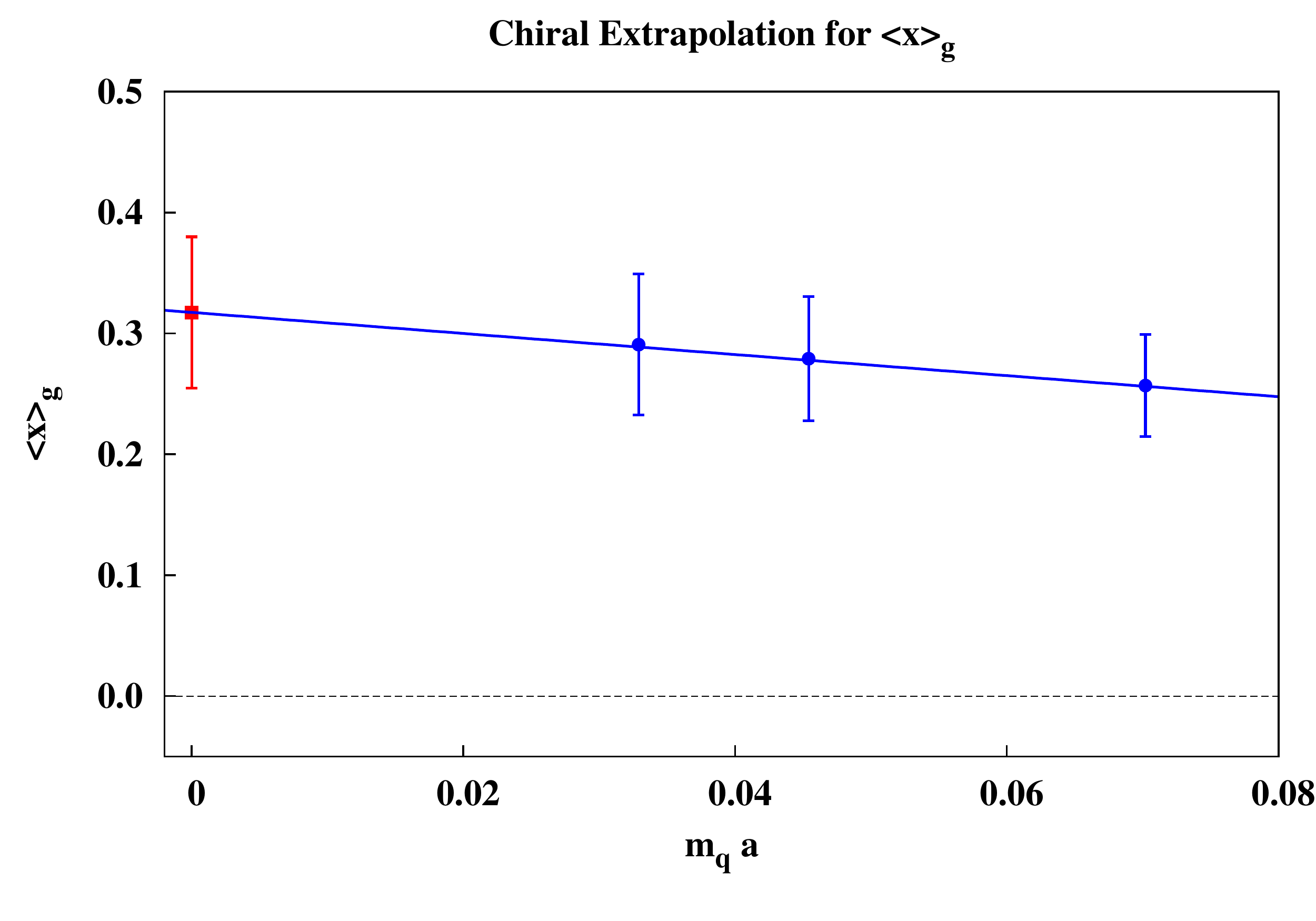}}
  \label{fig:chiral_x_1555_g}}
\caption{
 Plots for glue first moment:\ 
(a) ratio between three-point and two-point functions obtained by using 
Eq.~(\ref{eq:first_mom_rat_2}) at $\kappa_v = 0.1555$,\
and
(b) chiral extrapolation.}
\label{fig:glue_first_mom}
\end{figure}
\begin{figure}[htbp]
\centering
\subfigure[]
{\rotatebox{270}{\includegraphics[width=0.34\hsize]{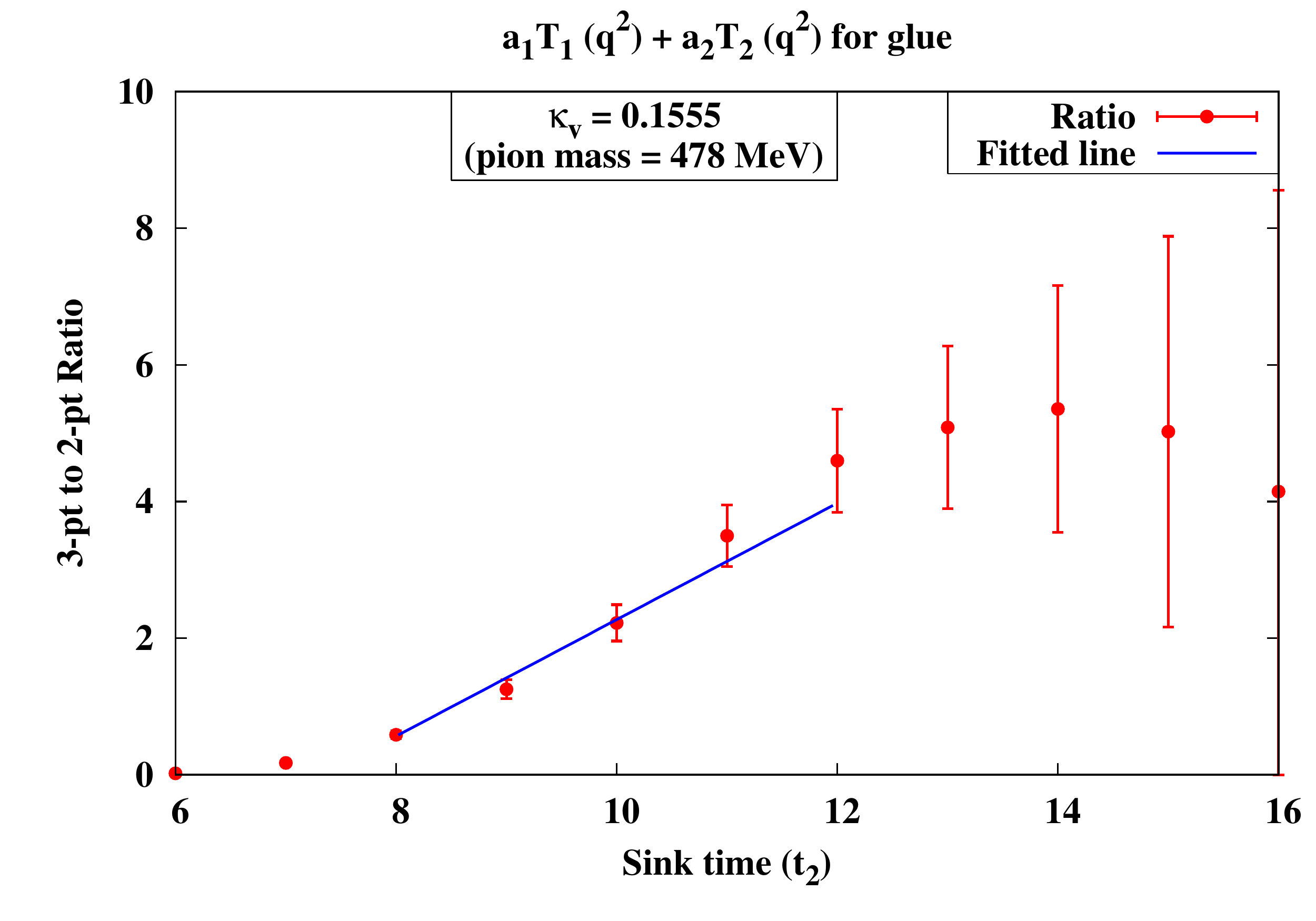}}
  \label{fig:slope_t1t2t3_1555_g}}
\subfigure[]
{\rotatebox{270}{\includegraphics[width=0.34\hsize]{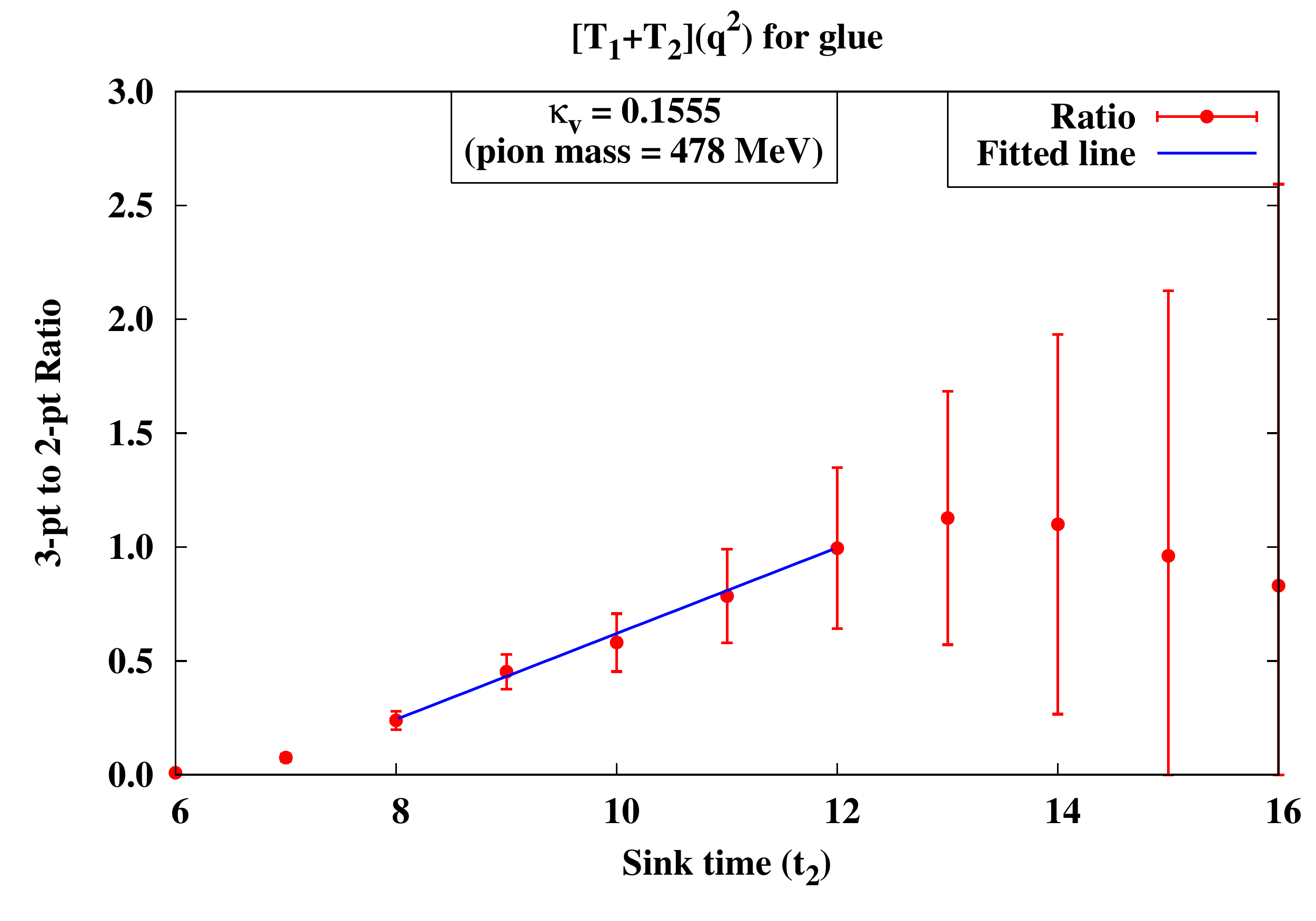}}
  \label{fig:slope_1555_g}}
\subfigure[]
{\rotatebox{270}{\includegraphics[width=0.34\hsize]{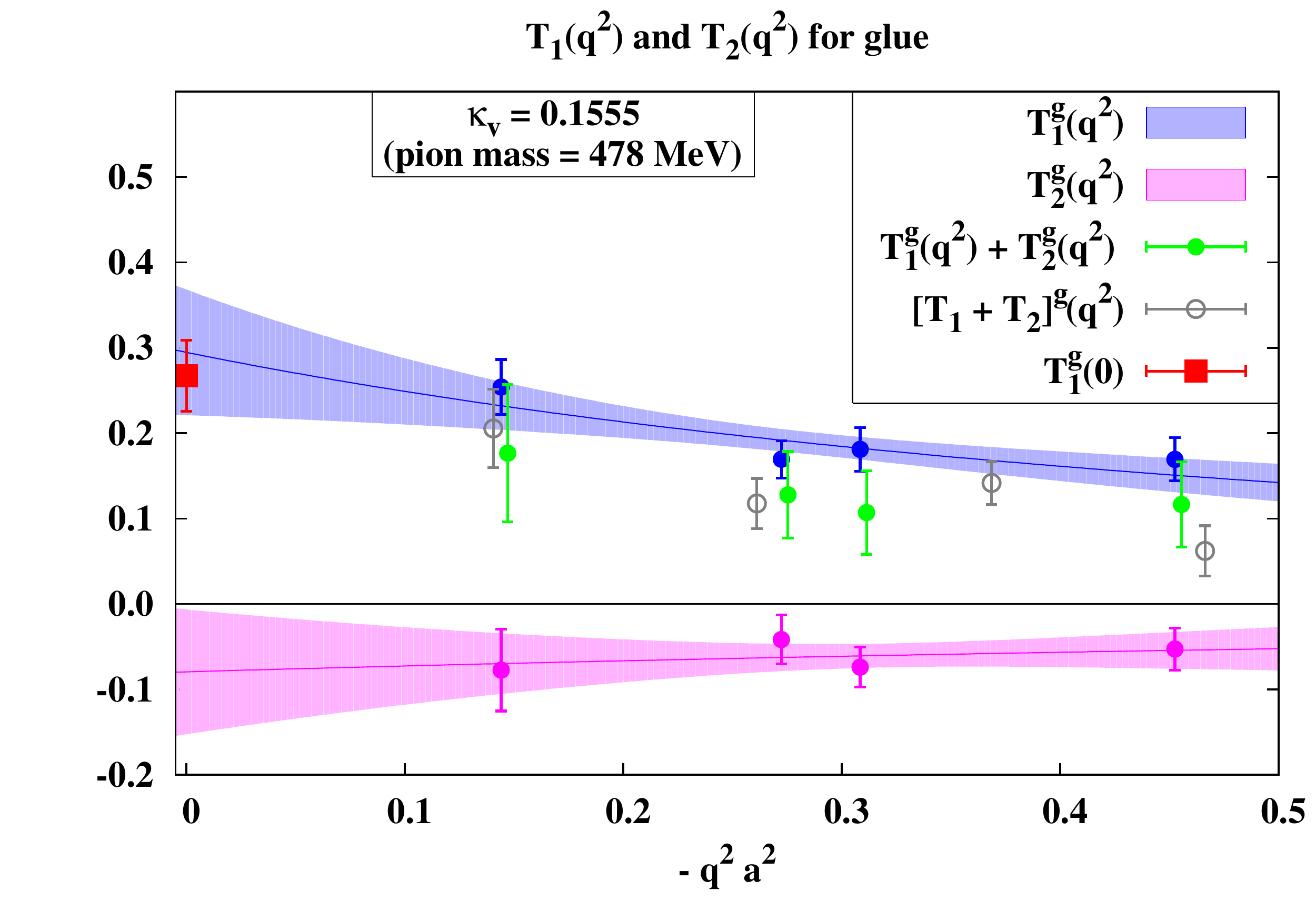}}
  \label{fig:T1T2_g}}
\subfigure[]
{\rotatebox{270}{\includegraphics[width=0.34\hsize]{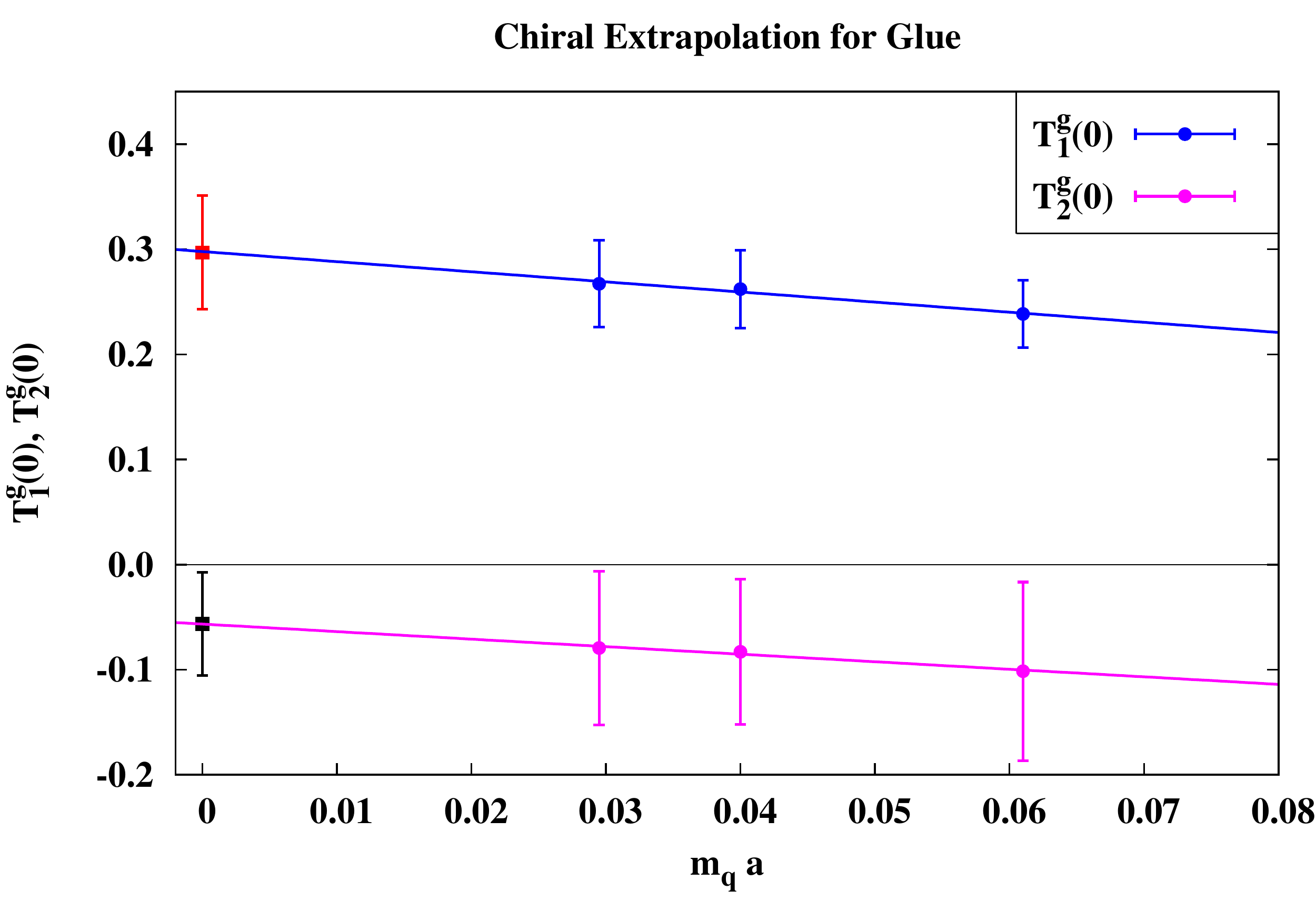}}
 \label{fig:chiral_fit_glue_ang_mom}}
\caption{
 Similar types of plots as in Fig.~\protect\ref{fig:DI} for the glue at  
 $\kappa_v = 0.1555$.}
\label{fig:glue}
\end{figure}

We perform the similar analysis for the glue momentum and angular momentum.\ The plots 
for glue first moment are shown in Figs.~\ref{fig:slope_x_1555_g} and 
\ref{fig:chiral_x_1555_g}.\ For angular momentum,\ they are plotted in 
Figs.~\ref{fig:slope_t1t2t3_1555_g},~\ref{fig:slope_1555_g},~\ref{fig:T1T2_g}~and~\ref{fig:chiral_fit_glue_ang_mom}.\ 
The first clear signal for the glue momentum fraction was seen with the 
overlap operator~\cite{Doi:2008hp}.\ Recently,\ the glue momentum fraction was calculated 
by using the Feynman-Hellmann theorem~\cite{Horsley:2012pz}.\ In our current work,\
clear signals of both the glue momentum and angular momentum fractions have been observed 
with direct calculation of the glue operators in the nucleon. 


\

In Table~\ref{tab:unren},\ we list the lattice results on the quark momentum fractions 
$\langle x\rangle \equiv T_1(0)$ for CI ($u$ and $d$) and DI ($u/d$ and $s$) as well as 
that for glue.\ We also list the corresponding $T_2(0)$ and total angular momenta fraction 
$2J = T_1(0) + T_2(0)$ for each quark flavor and glue.\ As explained in 
Sec.~\ref{subsec:results_ci},\ the $T_2(0)$ at $q^2 =0$ for CI($u$) and CI($d$) are obtained 
from separate dipole fits in $q^2$ as shown in Fig.~\ref{fig:quark_ang_mom_CT_1} while 
the $T_2(0)$ for CI($u+d$) is obtained from the dipole fit of the sum of CI($u$) and CI($d$) 
that leads to a smaller error than that obtained from the separate dipole fits.\ We note 
that the $T_2(0)$ from the quark and the glue sectors have similar magnitude but with opposite 
sign that results in cancellation within errors.\ This is consistent with Eq.~(\ref{eq:T_2_sum}) 
which results from momentum and angular momentum conservation.\ Consequently,\ the total 
unrenormalized momentum, $\langle x \rangle^q + \langle x \rangle^g = 0.95(7)$,\ and angular 
momentum,\ $2J^q + 2J^g = 0.95(9)$,\ are the same within errors and consistent with and 
close to unity.

\begin{table}[htbp]
  \centering
  \renewcommand{\arraystretch}{1.4}
  \begin{tabular}{|c||cc|cccc|c|}
    \hline\hline
    & {\bf CI(u)} & {\bf CI(d)}  & {\bf CI(u+d)} &  {\bf DI(u/d)} & {\bf DI(s)} & {\bf Glue} 
    & {\bf Total}\\
    \hline
    {\boldmath $\langle x \rangle$}
    & 0.408(38)  &  0.149(19) & 0.558(43) & 0.036(7) & 0.023(6) & 0.298(53) & 0.95(7) \\
    \hline
    {\boldmath $T_2(0)$} 
    & 0.283(107)  & -0.217(76) & 0.061(20) & -0.002(2) & -0.001(3) & -0.056(49) & 0.00(6)\\
    \hline
    {\boldmath $2J$} 
    &  0.691(122)  & -0.069(78) & 0.620(48) & 0.034(7) & 0.022(7) & 0.242(73) & 0.95(9)\\
    \hline\hline
  \end{tabular}
  \caption{Unrenormalized lattice results of quark and glue momenta and angular momenta.}
  \label{tab:unren}
  \end{table}

\subsection{Renormalization}
\label{sec:renormalization}

Before presenting the final results,\ we discuss renormalization and mixing of quark and glue 
operators and matching to $\overline{MS}$ scheme at a certain scale.\ The momenta 
$\langle x\rangle$ and angular momenta $J$ for the quarks and glue are calculated with 
lattice regularization.\ To match to the $\overline{MS}$ scheme at a scale $\mu$ in order to 
be able to compare with experiments,\ the renormalized matching and mixing of the 
momentum fraction (and angular momentum) can be written in the following matrix equation
\eqarray{
  \left[
    \begin{array}{c}
      \langle x\rangle^{\overline{MS}}_q (\mu, \mbox{CI}) \\
      \langle x\rangle^{\overline{MS}}_q (\mu, \mbox{DI}) \\
      \langle x\rangle^{\overline{MS}}_g (\mu)
    \end{array}
  \right] 
  &=& \left[
    \begin{array}{ccc} 
      Z_{qq} (a\mu, g_0) & 0  & 0 \\
      0    & Z_{qq} (a\mu, g_0) & Z_{qg} (a\mu, g_0) \\
      Z_{gq} (a\mu, g_0) &  Z_{gq} (a\mu, g_0)  & Z_{gg} (a\mu, g_0) 
    \end{array}
  \right]\,
  \left[
    \begin{array}{c}
       \langle x\rangle^{L}_q (\mbox{CI})  \\
       \langle x\rangle^{L}_q (\mbox{DI})  \\
       \langle x\rangle^{L}_g
    \end{array}
  \right],
\label{renorm}        
}
where the $\langle x\rangle^{L}_q$ and $\langle x\rangle^{L}_g$ are lattice matrix elements 
which satisfy the momentum sum rule,\ and the subscript ``q'' refers to the flavor-singlet 
quark component.\ The CI part corresponds to the moment of the parton distribution function 
for the valence and connected-sea (CS) quarks,\ whereas DI part is the corresponding moment 
for the disconnected-sea~\cite{Liu:1999ak}.\ The valence,\ CS and DS parton degrees of freedom 
are defined in the path-integral formulation of the hadronic tensor~\cite{Liu:1999ak},\ and the 
separation of CS from DS patrons has been achieved~\cite{Liu:2012ch} by combining HERMES 
data on the strangeness distribution~\cite{Airapetian:2008qf},\ the CT10 globally fitted parton 
distribution functions and the lattice calculation of the ratio of $\langle x\rangle$ of the strange 
to that of $u$ (or $d$) in the DI~\cite{Doi:2008hp}.\ It is important to note that valence and CS 
parton moments do not have contributions from the glue moment.\ Only the DS patron moment 
receives contributions from the glue moments through mixing.\ Since the energy-momentum 
tensors for the quark and glue are gauge invariant operators,\ their matrix elements do not mix 
with those of gauge variant operators~\cite{Joglekar:1975nu}.

\medskip

\noindent
The quark and glue momentum fractions in the $\overline{MS}$ scheme sum to unity 
provided the scheme-dependent renormalization constants,\ $Z(a\mu, g_0)$'s,\ satisfy 
the following constraints~\cite{Joglekar:1975nu,Ji:1995sv}
\eqarray{
Z_{qq} + Z_{gq} &=& 1\, ,\,  
Z_{qg} + Z_{gg}\, =\, 1 ,
\label{ren_constraint}
}
and the lattice quark and glue momentum fractions are normalized to satisfy the
momentum sum rule,\ i.e.
\begin{equation}
  \langle x\rangle^{L}_q + \langle x\rangle^{L}_g \, =\, 1 .
\label{lattice_norm_sum_rule}
\end{equation}
where $\langle x\rangle^{L}_q = \langle x\rangle^{L}_q (\mbox{CI}) 
+ \langle x\rangle^{L}_q (\mbox{DI})$.
The sum-rule improved lattice matrix elements in Eq.~(\ref{lattice_norm_sum_rule}) are 
defined as
\eq{
 \langle x\rangle^{L}_{q,g} = Z_{q,g}^L \langle x\rangle^L_{q,g} ,
}
where $\langle x\rangle^L_{q,g}$ are the unrenormalized matrix elements from the lattice 
calculation and $Z_{q,g}^L$ are the lattice normalization constants that account for lattice 
systematics.

\medskip

\noindent
Since both the momenta and angular momenta are derived from the same energy-momentum 
tensor operators,\ both $Z_q^L$ and $Z_g^L$ can be determined from the momentum and 
angular momentum sum rules
\eqarray{
  && Z_q^L  \langle x\rangle_q^L +  Z_g^L  
  \langle x\rangle_g^L = 1\, , 
  \label{lat_mom_sum}\\
  && Z_q^L  J_q^L + Z_g^L  J_g^L\, =\, \frac{1}{2} .
  \label{lattice_sum}
}
Even though the lattice calculated momenta and angular momenta are correlated, the direct 
fitting of $Z_{q,g}^L$ can lead to large errors since the values of $\langle x\rangle_q^L$ and 
$2J_q^L$ are close,\ as are those of $\langle x\rangle_g^L$ and $2J_g^L$.\ The condition number
of the $2 \times 2$ matrix of these matrix elements is $\sim 17$.\ Instead,\ one can choose to fit
$Z_{q,g}^L$ from the momentum sum rule in Eq.~(\ref{lat_mom_sum}) and 
\eq{
Z_q^L T_{2,q}^L (0)+ Z_g^L  T_{2,g}^L(0) = 0,
}
which leads to a smaller condition number of $\sim 8.6$,\ but the uncertainties in the lattice 
normalization factors $Z_{q,g}^L$ can still be large.

\medskip

\noindent
In view of the fact that the total unrenormalized lattice momentum 
$\langle x \rangle^q + \langle x \rangle^g = 0.95(7)$ and angular momentum 
$2J^q + 2J^g = 0.95(9)$ are the same within errors, we shall simply scale both to unity with 
$Z_q^L = Z_g^L =1.05$ and ignore their errors in this work.

\medskip

\noindent
For the renormalization constants,\ $Z_{qq}$,\ $Z_{qg}$,\ $Z_{gq}$,\ $Z_{gg}$ in 
Eq.~(\ref{renorm}),\ we shall compute them perturbatively.\ Lattice perturbation calculation 
has been carried out to match the energy-momentum tensor operators from the lattice to the 
$\overline{MS}$ scheme~\cite{Glatzmaier:2014sya}.\ To one-loop order,\ they are
\eqarray{
  Z_{qq} 
  &=& 1 + \frac{g_0^2}{16\pi^2}\, C_F\left(\frac{8}{3}\log(a^2\mu^2) + f_{qq}\right)\, , \,
  Z_{qg}
  \, =\, -\, \frac{g_0^2}{16\pi^2}\left(\frac{2}{3}\, N_f \log(a^2\mu^2) + f_{qg}\right)\, ,  
  \nonumber\\
  Z_{gq} 
  &=& -\, \frac{g_0^2}{16\pi^2}\, C_F\left(\frac{8}{3}\log(a^2\mu^2) + f_{gq}\right)\, , \, 
  Z_{gg}\, =\, 1 + \frac{g_0^2}{16\pi^2}\left(\frac{2}{3}\, N_f\log(a^2\mu^2) + f_{gg}\right) .
  \label{Z-factor}
}
For the negative mass parameter $\rho= 1.368$ used in the overlap operator,\ we obtain
$f_{qq} = -\, 7.60930$,\ $f_{gq} = -\, 2.37600$,\ $f_{qg} = 0.0$ and $f_{gg} = -\, 3.76900$.\ The 
details of the calculation are presented in Ref.~\cite{Glatzmaier:2014sya}.

\medskip

\noindent
We note that if we do not use the sum rule constraints for the lattice results,\ i.e. if we set 
$Z_{q,g}^L =1$,\ we find the total momentum fraction to be $0.92(7)$ and two times the total 
angular momentum fraction to be $0.92(9)$ in the $\overline{MS}$ scheme at $\mu =2$~GeV 
through Eq.~(\ref{Z-factor}).

\medskip

\noindent
We see that while the scheme- and scale-independent factors associated with the anomalous
dimensions $\gamma_{ij}$ together with the unity in the diagonal terms in Eq.~(\ref{renorm}) 
satisfy the constraints in Eq.~(\ref{ren_constraint}),\ the scheme-dependent finite factors 
$f_{ij}$ do not.\ This may be attributed to the artifact in the off-shell calculation of 
renormalization factors~\cite{Collins:1994ee}.\ In the literature,\ the finite factors $f_{qq}$ 
and $f_{qg}$ have been calculated to determine $Z_{qq}$ and $Z_{qg}$.\ On the other hand,\ 
$Z_{gq}$ and $Z_{gg}$ are simply defined from the constraints in 
Eq.~(\ref{ren_constraint})~\cite{Meyer:2007tm,Horsley:2012pz,Alexandrou:2013tfa} as
\eqarray{
Z_{gq} &=& 1 - Z_{qq}\, ,\,
Z_{gg}\, =\, 1 - Z_{qg} .
\label{Z-constraint}
}
Since we have calculated all the finite factors $f_{ij}$,\ we shall consider the average of the 
procedure such as the one in Eq.~(\ref{Z-constraint}) and replace $f_{ij}$ in 
Eq.~(\ref{Z-factor}) with $\tilde{f}_{ij}$ given by 
\eqarray{
  \tilde{f}_{qq} &=& \tilde{f}_{gq} \,=\, \frac{1}{2} (f_{qq} + f_{gq})\, ,\, 
  \tilde{f}_{qg}\, =\, \tilde{f}_{gg}\, =\, \frac{1}{2} (f_{qg} + f_{gg}) ,
  \label{shift-f}
}
so that the constraints in Eq.~(\ref{ren_constraint}) are satisfied.\ Although this procedure 
has an ambiguity,\ this systematic is expected to make negligible contributions to the final 
momentum and angular momentum fractions in the $\overline{MS}$ scheme. Since the 
prefactor $g_0^2/(16\pi^2) = 6.33 \times 10^{-3}$ is small, the effects in the finite factors 
in the renormalization constants are much smaller than unity.\ We find that the corresponding 
differences in the quark and glue momentum fractions due to the finite factors that are obtained 
by using Eqs.~(\ref{Z-constraint}) and (\ref{shift-f}) are less than $1$\% which is much smaller 
than the statistical errors of the physical quantities we calculate.

\medskip

\noindent
Since our inverse lattice spacing is determined to be $1/a = 1.74$~GeV from the nucleon
mass~\cite{Mathur:1999uf},\ $\log (a^2 \mu^2) = 0.279$ is small because $1/a$ is close 
to the scale $\mu \simeq 2$ GeV.\ Moreover,\ the factor $g_0^2/(16\pi^2) = 6.33 \times 10^{-3}$
is also small.\ As a result, the diagonal renormalization coefficients \ $Z_{qq} = 0.9641$ and
$Z_{gg} = 0.9881$ (for the quenched case with $N_f =0$) are close to unity and the off-diagonal
mixing coefficients\ $Z_{gq} = 0.0359$ and $Z_{qg} = 0.0119$ are close to zero.\ We see from 
Eq.~(\ref{renorm}) that there are only sub-percent changes from the lattice results to those in 
the $\overline{MS}$ scheme at $\mu \simeq 2$ GeV.\ We report our results in the 
$\overline{MS}$ scheme at  $\mu = 2$ GeV.

\subsection{Discussion}
\label{sec:discussion}

In Table~\ref{tab:chiral},\ we list the renormalized quark momentum fractions 
$\langle x\rangle \equiv T_1(0)$ for CI ($u$ and $d$) and DI ($u/d$ and $s$) as well as 
that of glue.\ We also list the corresponding $T_2(0)$ and total angular momenta fraction 
$2J = T_1(0) + T_2(0)$ for each quark flavor and glue.\ These values are obtained at 
$\mu = 2$~GeV in $\overline{MS}$ scheme as explained in Sec.~\ref{sec:renormalization}.\ 
To obtain results for different flavor,\ we note that $\langle  x\rangle_q^L (CI)$ is the linear
sum of those of $u$ and $d$ in the CI,\ and  $\langle  x\rangle_q^L (DI)$ is the linear
sum of those of $u, d$ and $s$ in the DI.\ Thus,\ in practice,\ Eq.~(\ref{renorm}) is extended 
to the bases of the direct product of flavor and CI and DI plus the glue,\ and the renormalization 
constants in Eq.~(\ref{Z-factor}) modified in such a way that $N_F$ is replaced with unity
and $f_{qg}$ replaced with $1/N_F$.\ The exception to this change is $Z_{gg}$ where the $N_F$ 
factor is zero for the present quenched calculation. 

\medskip

\noindent
We see from Table~\ref{tab:chiral} that the strange momentum fraction 
$\langle x\rangle_s = 0.024(6)$ is in the range of uncertainty of $\langle x\rangle_s$ from 
the CTEQ fitting of the parton distribution function from experiments which is 
$0.018 < \langle x\rangle_s < 0.040$~\cite{Lai:2007np}.\ The glue momentum fraction of 
$0.334 (55)$ is smaller than,\ say,\ the CTEQ4M fit of $0.42$ at 
$\mu = 1.6$~GeV~\cite{Lai:1996mg},\ but only by $1.5 \sigma$.\ The smallness of our 
value of $\langle x\rangle_g$ in comparison to the experiment could be in part due to the 
fact that ours is a quenched calculation.\ We expect the glue momentum fraction to be larger 
than the present result when dynamical configurations with light fermion are used in the 
calculation.

\medskip

\noindent
From Figs.~\ref{fig:quark_ang_mom_CT_2} and \ref{fig:T1T2_g} and Table~\ref{tab:chiral},\ 
we find that $\left[T_2^u(0) + T_2^d(0)\right]$ (CI) is positive and $T_{2}^g (0)$ is negative,\ 
so that the total sum including the small $\left[T_2^u (0) + T_2^d (0) + T_2^s (0)\right]$ (DI) 
can be naturally constrained to be zero (See Eq.~(\ref{eq:T_2_sum})) with the normalization 
constants $Z_q^L = 1.05$ and $Z_g^L = 1.05$ close to unity.\ In analogy to $F_2(0)$,\ 
the anomalous magnetic moment of the nucleon,\ $T_2(0)$,\ is termed as anomalous 
gravitomagnetic moment and has been shown to vanish for composite systems by Brodsky 
{\it et al.}~\cite{Brodsky:2000ii}.\ As we explained in Sec.~\ref{formalism},\ the vanishing of 
the total $T_2(0)$ is the consequence of momentum and angular momentum conservation. 
\begin{table}[htbp]
  \centering
  \renewcommand{\arraystretch}{1.4}
  \begin{tabular}{|c||cc|cccc|}
    \hline\hline
    & {\bf CI(u)} & {\bf CI(d)}  & {\bf CI(u+d)} &  {\bf DI(u/d)} & {\bf DI(s)} & {\bf Glue} \\
    \hline
    {\boldmath $\langle x \rangle$}
    & 0.413(38)  &  0.150(19) & 0.565(43) & 0.038(7) & 0.024(6) & 0.334(55) \\
    \hline
    {\boldmath $T_2(0)$} 
    & 0.286(108)  & -0.220(77) & 0.062(21) & -0.002(2) & -0.001(3) & -0.056(51) \\
    \hline
    {\boldmath $2J$} 
    &  0.700(123)  & -0.069(79) & 0.628(49) & 0.036(7) & 0.023(7) & 0.278(75)\\
    \hline
    {\boldmath $g_A$}
    &  0.91(11)  & -0.30(12)   & 0.62(9)  &  -0.12(1)  &  -0.12(1) & \--- \\
    \hline
    {\boldmath $2L$}
    &  -0.21(16)    &  0.23(15)   &  0.01(10)  &  0.16(1)  &  0.14(1) & \--- \\
    \hline\hline
  \end{tabular}
  \caption{Renormalized values in $\overline{MS}$ scheme at $\mu = 2$~GeV.}
  \label{tab:chiral}
\end{table}
\begin{figure}[h]
  \centering
  \subfigure[]
  {\rotatebox{0}%
    {\includegraphics[width=0.99\textwidth]{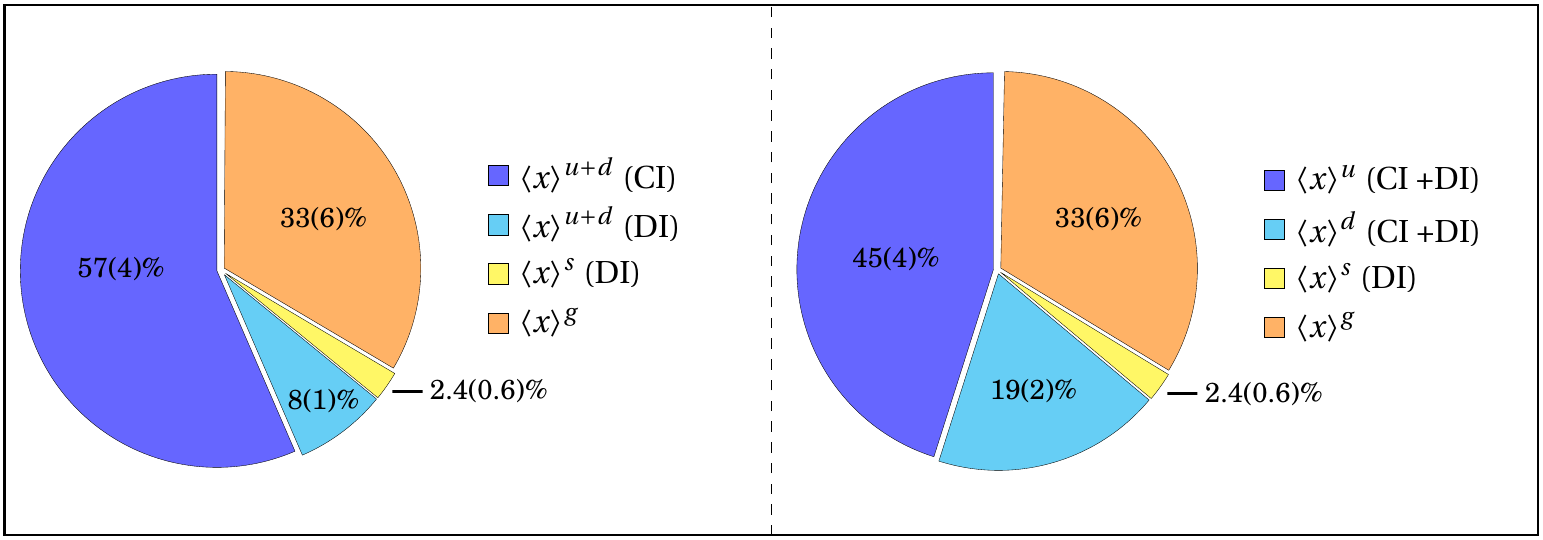}}
    \label{fig:pie_diag_first_mom}
  }
  \subfigure[]
  {\rotatebox{0}%
    {\includegraphics[width=0.99\textwidth]{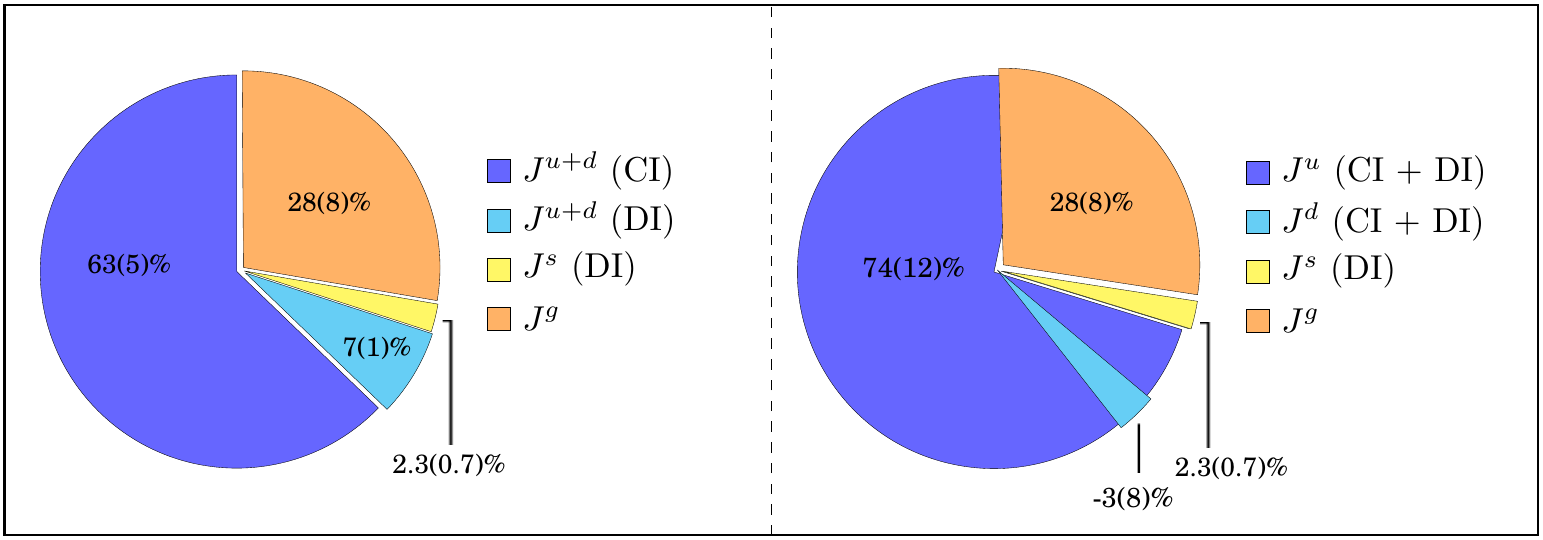}}
    \label{fig:pie_diag_am}
  }
  \subfigure[] 
  {\rotatebox{0}%
    {\includegraphics[width=0.99\textwidth]{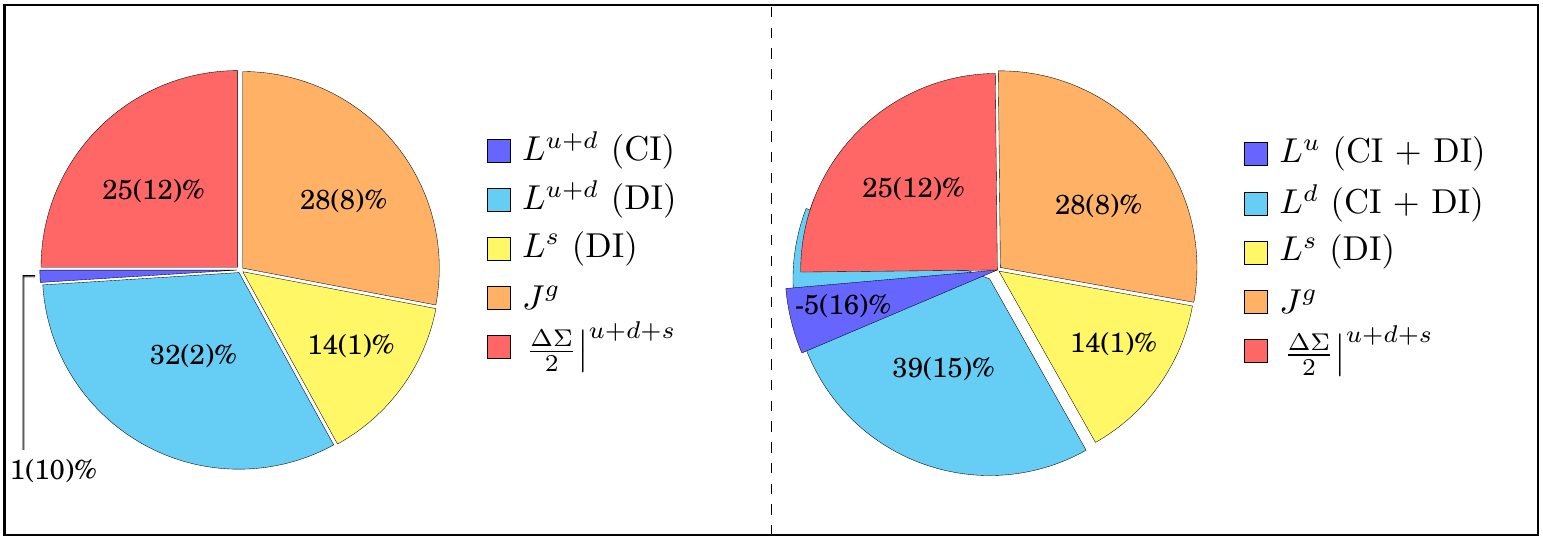}}
    \label{fig:pie_diag_orb_am}
  }
  \caption{Pie charts for the quark and gluon contributions to the\
    (a)\ momentum fraction,\ 
    (b)\ angular momenta,\ 
    and\
    (c)\ orbital angular momenta.\ 
    The left panels show the quark contributions separately for CI and DI,\ and the right panels 
    show the quark contributions for each flavor with CI and DI summed together for $u$ and 
    $d$ quarks.%
  }
  \label{fig:pie_diag}
\end{figure}

\medskip

\noindent
The flavor-singlet $g_A^0$ which is the quark spin contribution to the nucleon has been 
calculated before on the same lattice~\cite{Dong:1995rx}.\ We can subtract it from the 
total quark angular momentum fraction $2J$ to obtain the orbital angular momentum fraction 
$2L$ for the quarks.\ As we see in Table~\ref{tab:chiral},\ the orbital angular momentum 
fractions $2L$ for the $u$ and $d$ quarks in the CI have different signs and they add up to 
zero,\ i.e.\ $0.01(10)$.\ This is the same pattern seen with dynamical fermions configurations 
with light quarks~\cite{Hagler:2003jd,Brommel:2007sb,Bratt:2010jn,Alexandrou:2011nr,
Alexandrou:2013joa}.\ The large $2L$ for the $u/d$ and $s$ quarks in the DI is due to the 
fact that $g_A^0$ in the DI is large and negative,\ i.e.\ $-0.12(1)$ for each of the three 
flavors.\ All together,\ the quark orbital angular momentum constitutes a fraction of 
$0.47(13)$ of the nucleon spin.\ The majority of it comes from the DI.\ The quark spin fraction 
of the nucleon spin is $0.25(12)$ and glue angular momentum contributes a fraction of 
$0.28(8)$.\ We show all the different contributions to the momentum,\ angular momenta and 
orbital angular momenta in Figs.~\ref{fig:pie_diag_first_mom},\ \ref{fig:pie_diag_am} and 
\ref{fig:pie_diag_orb_am}.\ The left panels show the combinations of $u$ and $d$ contributions 
from CI and DI separately while the right panels show the contributions from the $u$ and 
$d$  quarks (with both CI and DI combined together).

\medskip

\noindent
We  note from Table~\ref{tab:chiral} that the orbital angular momenta contribution 
from each quark flavor is strongly dependent on the corresponding quark spin,\ particularly 
in the case of DI.\ As opposed to earlier calculations~\cite{Dong:1995rx,Fukugita:1994fh,
Gusken:1999as},\ the recent lattice calculations with light dynamical 
fermions~\cite{QCDSF:2011aa,Abdel-Rehim:2013wlz,Babich:2010at} have obtained smaller 
quark spin contributions from DI.\ However,\ preliminary study~\cite{Gong:2013} of the 
anomalous Ward identity with light valence overlap fermion on $2+1$-flavor dynamical 
domain wall fermion sea configurations suggests that the DI contributions are not small,
though this study has larger error bars.\ More detailed dynamical fermion calculations with 
controlled statistical and systematic errors are needed to settle this issue.  

\medskip

\noindent
We should point out that a small $\Delta u + \Delta d + \Delta s$ from the DI does not 
explain the small quark spin 
$g_A^0 = (\Delta u + \Delta d) (\rm{CI}) + (\Delta u + \Delta d + \Delta s) (\rm{DI}) 
\sim 0.25 $ from the global fitting of DIS~\cite{deFlorian:2009vb}, in view of the fact that 
most of the lattice calculation of $(\Delta u + \Delta d) (\rm{CI})$ is $\sim 0.6$ which is 
much larger than $0.25$.\ On the other hand,\ one could imagine that 
$(\Delta u + \Delta d) (\rm{CI})$ may turn out to be smaller than $0.6$ when the quark 
mass is close to the physical one in future lattice calculations,\ such as 
in~\cite{Alexandrou:2011nr} where $(\Delta u + \Delta d) (\rm{CI})$ is found to be much 
smaller than $0.6$ when the chiral extrapolation of the lattice results is carried out.\ However,\ 
this will not explain the octet 
$g_A^8 = (\Delta u + \Delta d) (\rm{CI}) + (\Delta u + \Delta d - 2 \Delta s) (\rm{DI})$.\ 
When both the CI and DI are small,\ the calculated $g_A^8$ will be smaller than the 
experimental value of $g_A^8 = 0.579(25)$~\cite{Close:1993mv}.\ Thus,\ it is difficult to 
explain simultaneously $g_A^0$ and $g_A^8$ with a small $(\Delta u + \Delta d + \Delta s)$ 
in DI.\ To clarify this issue,\ a full QCD simulation for $g_A^0$ and $g_A^8$ (both CI and DI) 
around the physical point by taking into account the $SU(3)$ breaking effect is necessary.

\medskip

\noindent
In the constituent quark model,\ the proton spin comes entirely from the quark spin.\ On the 
other hand,\ in the skyrmion,\ the total proton spin is from the collective rotational motion of 
the pion field~\cite{Adkins:1983ya}.\ What we find in the present calculation seems to suggest 
that the QCD picture,\ aside from the glue contribution,\ is somewhere in between these two 
models.\ Following Wilson's renormalization group approach to effective theories,\ it is 
suggested~\cite{Liu:1999kq} that the effective theory for baryons between the scale of 
$4\pi f_{\pi}$ and $\sim 300$ MeV may be a chiral quark model with renormalized couplings 
and renormalized meson,\ quark and gluon fields which preserve chiral symmetry.\ Models like 
the little bag model with skyrmion outside the MIT bag~\cite{Brown:1979ui},\ the cloudy bag 
model~\cite{Thomas:1981vc} and quark chiral soliton model~\cite{Wakamatsu:2006dy} could 
possibly delineate the pattern of division among the components of the proton spin with large 
quark orbital angular momentum contribution.


\section{Summary}
\label{sec:summary}
In summary,\ we have carried out a complete calculation of the quark and glue momentum 
and angular momentum in the nucleon for the first time on a quenched $16^3 \times 24$ 
lattice with three quark masses.\ The calculation includes both the connected insertion (CI) 
and disconnected insertion (DI) of the three-point functions for the quark energy-momentum 
tensor.\ We have used complex $Z_2$ noise to estimate the quark loops in the DI and the 
gauge field tensor from the overlap operator in the glue energy-momentum tensor.\ We find 
that reasonable signals can be obtained for the glue operator constructed from the overlap 
Dirac operator.\ After chiral extrapolation,\ the momentum and angular momentum sum rules 
are used to normalize the quark and glue momentum and angular momentum fractions on the 
lattice.\ The renormalization and mixing of the quark and glue energy-momentum operators 
are obtained through one-loop perturbation,\ and the final results are reported in the 
$\overline{MS}$ scheme at $2$~GeV.\ The renormalized momentum fractions for the quarks 
are $0.565(43)$ for the CI and $0.100(15)$ for the DI.\ The glue momentum fraction is 
$0.334(55)$.\ We have demonstrated that the vanishing anomalous gravitomagnetic moment 
(see Eq.~(\ref{eq:T_2_sum})) is a consequence of momentum and angular momentum 
conservation.

\medskip

\noindent
After subtracting the quark spin ($g_A^0$) from a previous calculation on the same 
lattice~\cite{Dong:1995rx} from the angular momentum $2J$,\ we obtain the orbital angular 
fraction $2L$.\ In the CI,\ we find that the $u$ quark contribution is negative,\ while the $d$ 
quark contribution is positive.\ The sum is $0.01(10)$ which is small.\ This behavior is the 
same as observed in dynamical calculation with light quarks~\cite{Hagler:2003jd,
  Brommel:2007sb,Bratt:2010jn,Alexandrou:2011nr,Alexandrou:2013joa}.\ The majority of the 
quark orbital angular momentum turns out to come from the DI,\ because the quark spin from 
the DI is large and negative for each of the three flavors.\ In the end,\ we find the quark orbital 
angular momentum,\ the quark spin,\ and glue angular momentum fractions of the nucleon spin 
are $0.47(13), 0.25(12)$ and $0.28(8)$,\ respectively.

\medskip

\noindent
Finally,\ this work should be extended to dynamical fermion calculations with light quarks and
continuum and large volume limits to control the systematic errors of lattice QCD.\ We are in 
the process of carrying out the same calculation with the valence overlap fermion on 
$2+1$-flavor dynamical domain wall fermion sea configurations to remove the systematic 
errors due to the quenched approximation.


\acknowledgments

This work is partially supported by U.S.\ DOE Grant No.\ DE-FG05-84ER40154 and the Center
for Computational Sciences of the University of Kentucky.\ The work of M.~Deka is partially
supported by the Institute of Mathematical Sciences,\ India.\ The work of T.~Doi is supported 
in part by MEXT Grant-in-Aid for Young Scientists (B) (24740146).\ We would like to thank 
Igor V.\ Anikin,\ Ying Chen,\ and Oleg V.\ Teryaev for useful discussions and comments.



\phantomsection
\addcontentsline{toc}{chapter}{Appendix}

\vskip 20pt

\appendix

\section{Solving System of Kinematical Equations}
\label{appsec:kine_eqs}

In this section,\ we will discuss how to solve a system of kinematical equations to extract 
$T_1$,\ $T_2$ and $T_3$.\ As mentioned in Sec.~\ref{subsec:ratioscorrelationfunctions}),\ 
we need to combine several kinematics into the ratios in 
Eq.~(\ref{eq:three_pt_2_two_pt_rat_1}) for CI or in 
Eq.~(\ref{eq:three_pt_2_two_pt_rat_2}) for DI at a particular $q^2$ in order to separate 
$T_1(q^2),\ T_2(q^2)$ and $T_3(q^2)$.\ Using the available momenta,\ we obtain several 
ratios (for both polarized and unpolarized nucleons) for all the three directions of the 
operator,\ ${\mathcal T}_{4i}$,\ at every $q^2$.\ From these ratios,\ one can set up 
kinematical equations to solve for $T_1$,\ $T_2$ and $T_3$.\ For simplicity,\ we will consider 
the CI only as we have considered $\vec p\,' = ( 1, 0, 0 )$ in this case in order to reduce 
computational cost.\ The procedure for DI will be similar except that we will have more 
available momenta.

\medskip

If we consider the lowest $q^2$ ($= 0.1460$ for $\kappa = 0.154$) and 
$\vec p\,' = ( 1, 0, 0 )$,\ we obtain the following five different equations as
\eqarray{
  \label{eq:kin_eq_ci_1}
  &  & \frac{1}{4}\,
  \bigg[R^{\unpol}_{41} (0,1,0) + R^{\unpol}_{41} (0,-1,0) + R^{\unpol}_{41} (0,0,1) 
  + R^{\unpol}_{41} (0,0,-1)\bigg]
  \nonumber\\
  &=& \frac{1}{4}\, \frac{1}{\sqrt{E_{p'} (E_{p'} + m)E_p (E_p + m)}}\nonumber\\
  &  & \bigg[T_1 (q^2)
    \Big\{p'_1\, (E_{p'} + E_p)\, (3\, E_{p'} + E_p + 4\, m)\Big\}\nonumber\\
  &+& \frac{1}{2m} T_2 (q^2)
    \Big\{p'_1\, (E_{p'} - E_p)^2\, (E_{p'} + E_p)
    -\, p'_1\, q_2^2\, (3 E_{p'} + E_p + 2 m)\Big\}\bigg] ,\\
  &  & \nonumber\\
  &  & \nonumber\\
  \label{eq:kin_eq_ci_2}
  &  & \frac{1}{4}\,
  \bigg[R^{\unpol}_{42} (0,1,0) - R^{\unpol}_{42} (0,-1,0) + R^{\unpol}_{43} (0,0,1) 
  - R^{\unpol}_{43} (0,0,-1)\bigg]
  \nonumber\\
  &=& \frac{1}{4}\, \frac{1}{\sqrt{E_{p'} (E_{p'} + m)E_p (E_p + m)}}\nonumber\\
  &  & \bigg[T_1 (q^2)
    \Big\{(- 2 q_2)\, (E_{p'} + m)\, (E_{p'} + E_p)\Big\} 
    + \frac{1}{2m} T_2 (q^2)
    \Big\{(- q_2)\, (E_{p'} + m)\, (E_{p'}^2 + E_p^2 - q_2^2)\Big\}\nonumber\\ 
  &+& \frac{2}{m} T_3 (q^2)
    \Big\{q_2\, (E_{p'} - E_p)\, (E_{p'} + m)\, (E_p - E_{p'} + 2m)\Big\}\bigg] ,\\ 
  &  & \nonumber\\
  &  & \nonumber\\
  \label{eq:kin_eq_ci_3}
  &  & \frac{1}{2}\,
  \bigg[R^{\pol}_{41} (0,1,0) - R^{\pol}_{41} (0,-1,0)\bigg]
  \nonumber\\
  &=& \frac{1}{4}\, \frac{1}{\sqrt{E_{p'} (E_{p'} + m)E_p (E_p + m)}}\nonumber\\
  &  & \bigg[T_1 (q^2)
    \Big\{(- q_2)\, ((E_{p'} + m)\, (E_{p'} + E_p) + 2 p_1'\,^2)\Big\}\nonumber\\ 
  &+& \frac{1}{2m} T_2 (q^2)
    \Big\{(- q_2)\, (E_{p'} + m)^2\, (E_{p'} + E_p)
    - p'_1\, q_2^2\, (3 E_{p'} + 3 E_p + 4m)\Big\}\bigg] ,\\ 
  &  & \nonumber\\
  &  & \nonumber\\  
  \label{eq:kin_eq_ci_4}
  &  & \frac{1}{2}\,
  \bigg[R^{\pol}_{42} (0,1,0) - R^{\pol}_{42} (0,-1,0)\bigg]
  \nonumber\\
  &=& \frac{1}{4}\, \frac{1}{\sqrt{E_{p'} (E_{p'} + m)E_p (E_p + m)}}\nonumber\\
  &  & \bigg[T_1 (q^2)
    \Big\{(- p'_1)\, (E_{p'}^2 - E_p^2 - q_2^2)\Big\} 
    + \frac{1}{2m} T_2 (q^2)\, 
    \Big\{(- p'_1)\, (E_{p'} + E_p + 2 m)\, (E_{p'}^2 - E_p^2 - q_2^2)\Big\}\nonumber\\ 
  &+& \frac{2}{m} T_3 (q^2)
    \Big\{p'_1\, q_2^2\, (E_{p'} - E_p)\Big\}\bigg] , \\
  &  & \nonumber\\
  &  & \nonumber\\
  \label{eq:kin_eq_ci_5}
  &  & \frac{1}{2}\,
  \bigg[R^{\pol}_{42} (0,0,1) - R^{\pol}_{42} (0,0,-1)\bigg]
  \nonumber\\
  &=& \frac{1}{4}\, \frac{1}{\sqrt{E_{p'} (E_{p'} + m)E_p (E_p + m)}}\nonumber\\
  &  & \bigg[T_1 (q^2)
    \Big\{(- p'_1)\, (E_{p'}^2 - E_p^2)\Big\} 
    + \frac{1}{2m} T_2 (q^2)\, 
    \Big\{(- p'_1)\, (E_{p'}^2 - E_p^2)\, (E_{p'} + E_p + 2m) \nonumber\\ 
  &  & + (-p'_1)\, (E_{p'} + E_p)\, q_3^2\Big\}\bigg] ,
}
where $R$'s are the ratios in Eq.~(\ref{eq:three_pt_2_two_pt_rat_1}),\ e.g.\ the 
notation $R^{\unpol}_{41} (0,1,0)$ signifies the ratio for the unpolarized three-point functions 
corresponding to the ${\mathcal T}_{41}$ operator with a momentum transfer of 
${\vec q} = (0,1,0)$.

\smallskip

\noindent
For convenience,\ we shall write the  Eqs.~(\ref{eq:kin_eq_ci_1}),\ (\ref{eq:kin_eq_ci_2}),\ 
(\ref{eq:kin_eq_ci_3}),\ (\ref{eq:kin_eq_ci_4}),\ (\ref{eq:kin_eq_ci_5}) in the following 
manner
\eqarray{
\label{eq:kin_eq_ci_1_1}
R_1 &=& a_{1,1}\, T_1 (q^2) + a_{2,1}\, T_2 (q^2) + a_{3,1}\, T_3 (q^2) , \\
\label{eq:kin_eq_ci_2_1}
R_2 &=& a_{1,2}\, T_1 (q^2) + a_{2,2}\, T_2 (q^2) + a_{3,2}\, T_3 (q^2) , \\
\label{eq:kin_eq_ci_3_1}
R_3 &=& a_{1,3}\, T_1 (q^2) + a_{2,3}\, T_2 (q^2) + a_{3,3}\, T_3 (q^2) , \\
\label{eq:kin_eq_ci_4_1}
R_4 &=& a_{1,4}\, T_1 (q^2) + a_{2,4}\, T_2 (q^2) + a_{3,4}\, T_3 (q^2) , \\
\label{eq:kin_eq_ci_5_1}
R_5 &=& a_{1,5}\, T_1 (q^2) + a_{2,5}\, T_2 (q^2) + a_{3,5}\, T_3 (q^2) .
}
Here,\ $a_{i,j}$'s are the constant coefficients of $T_1(q^2),\ T_2(q^2)$ and 
$T_3(q^2)$ which include the factor,\ 
$\displaystyle\frac{1}{4}\, \frac{1}{\sqrt{E_{p'} (E_{p'}+m)E_p(E_p+m)}}$.\ 
However,\ the Eqs.~(\ref{eq:kin_eq_ci_1_1}),\ (\ref{eq:kin_eq_ci_2_1}),\ 
(\ref{eq:kin_eq_ci_3_1}),\ (\ref{eq:kin_eq_ci_4_1}) and (\ref{eq:kin_eq_ci_5_1}),\
though different,\ are numerically correlated since they are computed on the same 
set of configurations.\ Such correlations are taken into account by constructing a 
covariance matrix,\ $C$,\ between these equations.\ This allows us to define the 
corresponding $\chi^2$ as
\eqarray{
  \chi^2 
  &=& \displaystyle\sum_{ij}^N 
  \Big[R_i - (a_{1,i}\, T_1 + a_{2,i}\, T_2 + a_{3,i} T_3)\Big]\, 
  C_{ij}^{-1}\, 
  \Big[R_j - (a_{1,j}\, T_1 + a_{2,j}\, T_2 + a_{3,j} T_3)\Big] ,
\label{eq:app:chi_sq_ratio}
}
where $N$  is the number of equations which is equal to $5$ in this case.\ We then 
solve for $T_1,\ T_2$ and $T_3$ by imposing the following minimization conditions on 
the $\chi^2$ obtained from Eq.~(\ref{eq:app:chi_sq_ratio}) as
\eqarray{
\frac{\partial\chi^2}{\partial T_1} &=& 0,\ 
\frac{\partial\chi^2}{\partial T_2}\, =\, 0,\
\frac{\partial\chi^2}{\partial T_3}\, =\, 0 .
\label{eq:app:min_chi_sq_ratio}
}
This results in the three following equations
\eqarray{
\label{eq:app:ind_eq_ratio_1}
R'_1 &=& a^1_1\, T_1 + a^1_2\, T_2 + a^1_3\, T_3 , \\
\label{eq:app:ind_eq_ratio_2}
R'_2 &=& a^2_1\, T_1 + a^2_2\, T_2 + a^2_3\, T_3 , \\
\label{eq:app:ind_eq_ratio_3}
R'_3 &=& a^3_1\, T_1 + a^3_2\, T_2 + a^3_3\, T_3 ,
}
where
\eqarray{
a^m_k &=& 2\, a_{m,i}\, C_{ij}^{-1}\, a_{k,j},\, \,
R'_k \,=\, 2\, a_{k,i}\, C_{ij}^{-1}\, R_j, \hspace{15mm}
(m, k = 1, 2, 3),
}
and the sum over $i, j$ is implied according to Einstein's summation rule.\ 
Solving Eqs.~(\ref{eq:app:ind_eq_ratio_1}),\ (\ref{eq:app:ind_eq_ratio_2}) and 
(\ref{eq:app:ind_eq_ratio_3}),\ we can separate $T_1$,\ $T_2$ and $T_3$ at 
that $q^2$.

\vspace{15mm}


\phantomsection

\end{document}